\documentclass[journal]{IEEEtran}
\usepackage{cite}
\usepackage{amsmath}
\usepackage{times}
\usepackage{epsfig}
\usepackage{graphicx}
\usepackage{amsmath}
\usepackage{amssymb}
\usepackage{tabularray}
\usepackage{algorithm}
\usepackage{algorithmic}
\usepackage{stfloats}

\usepackage{caption}  
\usepackage{longtable} 
\usepackage{float} 
\usepackage{url}
\usepackage{subcaption}
\usepackage{xurl}
\usepackage{afterpage}
\usepackage{comment}

\usepackage{makecell} 

\usepackage{booktabs} 
\usepackage{array}
\usepackage{longtable}
\usepackage[table]{xcolor}
\usepackage[utf8]{inputenc}
\usepackage[T1]{fontenc}
\usepackage{graphicx}
\usepackage{tabularx}
\usepackage{caption}
\usepackage{adjustbox}
\usepackage{supertabular} 
\usepackage{multirow}      
\usepackage{booktabs}      
\usepackage{lipsum}        
\usepackage{hyperref}  
\usepackage{url}       

\ifCLASSINFOpdf
\else
\fi

\begin{document}
%
\title{SDP: A Unified Protocol and Benchmarking Framework for Reproducible Wireless Sensing}
%
%

%
%


    \author{Di Zhang, Jiawei Huang,~\IEEEmembership{Student Member,~IEEE,}  Yuanhao Cui, Xiaowen Cao,~\IEEEmembership{Member,~IEEE,}
    \\ Tony Xiao Han, Xiaojun Jing,~\IEEEmembership{Member,~IEEE,} Christos Masouros,~\IEEEmembership{Fellow,~IEEE}
    \thanks{Di Zhang, Jiawei Huang, Yuanhao Cui, and Xiaojun Jing are with the School of Information and Communication Engineering, Beijing University of Posts and Telecommunications, Beijing 100876, China. (e-mail: amandazhang, jiawei.aki, yuanhao.cui, jxiaojun@bupt.edu.cn).}
    
    \thanks{Xiaowen Cao is with the College of Electronics and Information Engineering, Shenzhen University, Shenzhen 518060, China (e-mail: caoxwen@szu.edu.cn).}
    
    \thanks{Tony Xiao Han is with the Huawei Technology Co. Ltd., Guangdong 518129, China (e-mail: tony.hanxiao@huawei.com).}
    
    \thanks{Christos Masouros is with the Department of Electronic and Electrical Engineering, University College London, Torrington Place, London, WC1E7JE, UK (email: c.masouros@ucl.ac.uk)}

}

\markboth{Journal of \LaTeX\ Class Files,~Vol.~14, No.~8, August~2015}%
{Shell \MakeLowercase{\textit{et al.}}: Bare Demo of IEEEtran.cls for IEEE Journals}

%



\maketitle

\begin{abstract}
Learning-based wireless sensing has made rapid progress, yet the field still lacks a unified and reproducible experimental foundation.
Unlike computer vision, wireless sensing relies on hardware-dependent channel measurements whose representations, preprocessing pipelines, and evaluation protocols vary significantly across devices and datasets, hindering fair comparison and reproducibility.

This paper proposes the Sensing Data Protocol (SDP), a protocol-level abstraction and unified benchmark for scalable wireless sensing. SDP acts as a standardization layer that decouples learning tasks from hardware heterogeneity. To this end, SDP enforces deterministic physical-layer sanitization, canonical tensor construction, and standardized training and evaluation procedures, decoupling learning performance from hardware-specific artifacts. Rather than introducing task-specific models, SDP establishes a principled protocol foundation for fair evaluation across diverse sensing tasks and platforms. Extensive experiments demonstrate that SDP achieves competitive accuracy while substantially improving stability, reducing inter-seed performance variance by orders of magnitude on complex activity recognition tasks.
A real-world experiment using commercial off-the-shelf Wi-Fi hardware further illustrating the protocol’s interoperability across heterogeneous hardware. By providing a unified protocol and benchmark, SDP enables reproducible and comparable wireless sensing research and supports the transition from ad hoc experimentation toward reliable engineering practice.
\end{abstract}


%
\IEEEpeerreviewmaketitle

\section{Introduction}\label{section1}

The evolution toward sixth-generation (6G) networks is driving a fundamental transformation of wireless systems from communication-centric infrastructures to Integrated Sensing and Communications (ISAC) platforms~\cite{9606831,10908560, zhang2025integratedsensingcommunicationsyears}. By exploiting the rich electromagnetic spectrum and large-scale antenna arrays, wireless networks are increasingly expected to sense human activities, device presence, and environmental dynamics, enabling applications such as device-free recognition, localization, tracking, and vital sign monitoring~\cite{10387517,9737357,10969844}. This new paradigm is becoming essential for creating intelligent environments, as it offers advantages in robustness to different lighting conditions and more stringent privacy requirements compared to vision-based methods~\cite{miao2025wi}.

Recent years have witnessed rapid progress in learning-based wireless sensing~\cite{fu2024wi,8514811,strohmayer2024data}, where deep neural networks extract latent representations from physical-layer measurements, particularly Channel State Information (CSI). CSI provides fine-grained channel responses that capture both static multipath structures and dynamic perturbations induced by human motion~\cite{wang2025surveywifisensinggeneralizability}. However, a fundamental domain wall persists between high-level learning models and physical-layer realities. Unlike computer vision, where raw data follows a canonical, hardware-agnostic RGB pixel-grid abstraction, wireless sensing lacks a unified representation. CSI measurements are inherently coupled with hardware-specific imperfections (e.g., carrier frequency offsets, sampling jitter) and vendor-dependent implementations (e.g., subcarrier layouts)~\cite{10.1145/3631429}. Consequently, models trained on one device (e.g., Intel 5300) often suffer severe performance degradation or require complete architectural redesign when transferred to another (e.g., Atheros), limiting generalization and reproducibility.

This device heterogeneity has created a fragmented research landscape. Pioneering benchmarks such as Widar3.0~\cite{9516988} and DeepSense 6G~\cite{alkhateeb2023deepsense} have significantly advanced the field by providing large-scale measurements for Wi-Fi gesture recognition and multimodal ISAC sensing, respectively. However, these contributions primarily validate sensing under specific or isolated settings. While recent studies, such as WiMANS~\cite{huang2024wimansbenchmarkdatasetwifibased}, have begun to address the complexities of multi-user scenarios, sophisticated multi-modal benchmarks like MM-Fi~\cite{yang2023mm} and XRF V2~\cite{10.1145/3749521} often report that Wi-Fi sensing yields inferior performance compared to LiDAR or wearable IMUs. We argue that this performance gap is, at least in part, attributable to the absence of a unified protocol. Without such a protocol, algorithmic improvements are frequently confounded by inconsistent preprocessing choices and hardware-dependent distortions, rendering fair, reproducible, and cross-domain comparisons fundamentally challenging.

To bridge this gap, this paper proposes the Sensing Data Protocol (SDP), a unified framework that systematizes the entire wireless sensing pipeline—from signal sanitization to evaluation. Specifically, we establish a deterministic preprocessing protocol to rigorously mitigate hardware impairments (e.g., sampling time offset (STO) and carrier frequency offset (CFO)). By adopting a protocol-driven Canonical Mapping strategy, SDP projects raw heterogeneous signals into a standardized Canonical CSI Tensor ($\mathcal{X} \in \mathbb{C}^{A \times K \times T}$) via physically motivated interpolation, effectively normalizing device-specific subcarrier layouts into a device-agnostic space. Complementing this protocol, we instantiate the benchmark with a fixed Transformer backbone as a transparent probe, not a task-optimized model. This allows for rigorous, reproducible comparisons across diverse sensing tasks without the need for task-specific architectural engineering.

\begin{table*}
\centering
\caption{Comparison of SDP with Representative Wireless Sensing Datasets}
\label{table-Related Datasets}
\begin{tblr}{
  width = \linewidth,
  colspec = {Q[90]Q[155]Q[105]Q[50]Q[50]Q[50]Q[50]Q[50]Q[50]Q[50]},
  cells = {c},
  cell{1}{1} = {r=2}{},
  cell{1}{2} = {r=2}{},
  cell{1}{3} = {r=2}{},
  cell{1}{4} = {c=3}{},
  cell{1}{7} = {c=4}{},
  vlines,
  hline{1,3-11} = {-}{},
  hline{2} = {4-10}{},
}
\textbf{Dataset}        & \textbf{Scenarios}                               & \textbf{Scale}            & \textbf{Modalities} &                 &                 & \textbf{Sensing Tasks} &                     &                  &                    \\
                        &                                                  &                           & \textbf{Wi-Fi}      & \textbf{mmWave} & \textbf{Vision} & \textbf{Loc./ Track}   & \textbf{Act./ Pose} & \textbf{Imaging} & \textbf{Vital/ ID} \\
\textbf{SDP}\cite{sdp8}~            & In/Outdoor                                       & 400~h, 14.1~TB            & $\bullet$           & $\bullet$       & $\bullet$       & $\bullet$              & $\bullet$           & $\bullet$        & $\bullet$          \\
\textbf{ImgFi}\cite{10190332}~         & Indoor                                           & -                         & $\bullet$           & $-$             & $-$             & $-$                    & $\bullet$           & $-$              & $-$                \\
\textbf{WiMANS}\cite{huang2024wimansbenchmarkdatasetwifibased}~        & Indoor                                           & {9.4~h, \\11286 samples~} & $\bullet$           & $-$             & $\bullet$       & $\bullet$              & $\bullet$           & $-$              & $-$                \\
\textbf{MM-Fi}\cite{yang2023mm}~         & Indoor                                           & {1080 samples, \\320 k frames  }             & $\bullet$           & $\bullet$       & $\bullet$       & $-$                    & $\bullet$           & $-$              & $-$                \\
\textbf{XRF V2}\cite{10.1145/3749521}~        & Dining, Study, Bedroom & 16~h                      & $\bullet$           & $-$             & $-$             & $-$                    & $\bullet$           & $-$              & $-$                \\
\textbf{EyeFi}\cite{9183685}~          & Lab, Kitchen  & 13h                       & $\bullet$           & $-$             & $\bullet$       & $\bullet$              & $-$                 & $-$              & $\bullet$          \\
\textbf{mmWave Gesture}\cite{yan2023mmgesture} & Home, Office, Lab               & {10~h, \\24050 samples} & $-$                 & $\bullet$       & $-$             & $-$                    & $\bullet$           & $-$              & $-$                \\
\textbf{DeepSense 6G}\cite{alkhateeb2023deepsense}   & In/Outdoor (Day/Night)                           & 1 M+ samples             & $\bullet$           & $\bullet$       & $\bullet$       & $\bullet$              & $-$                 & $-$              & $-$                
\end{tblr}
\end{table*}

The main contributions of this work are summarized as follows.
\begin{itemize}
    \item We propose the SDP, which introduces a Canonical CSI Tensor abstraction to represent heterogeneous CSI measurements under a unified mathematical interface. This canonical representation decouples downstream learning algorithms from device-specific subcarrier layouts and frontend implementations, enabling cross-device and cross-dataset evaluation in a standardized form.
    
    \item We develop a reproducible preprocessing workflow that deterministically mitigates dominant hardware impairments such as STO and CFO, and maps heterogeneous CSI measurements into a standardized Canonical CSI Tensor through protocol-driven frequency projection and windowing. This design reduces ambiguity in preprocessing choices and avoids dataset-dependent signal processing heuristics, thereby providing a consistent and stable input space for benchmarking.

\begin{figure*}[htp!]
    \centering
    \captionsetup{justification=raggedright}    
    \includegraphics[width=0.95\linewidth]{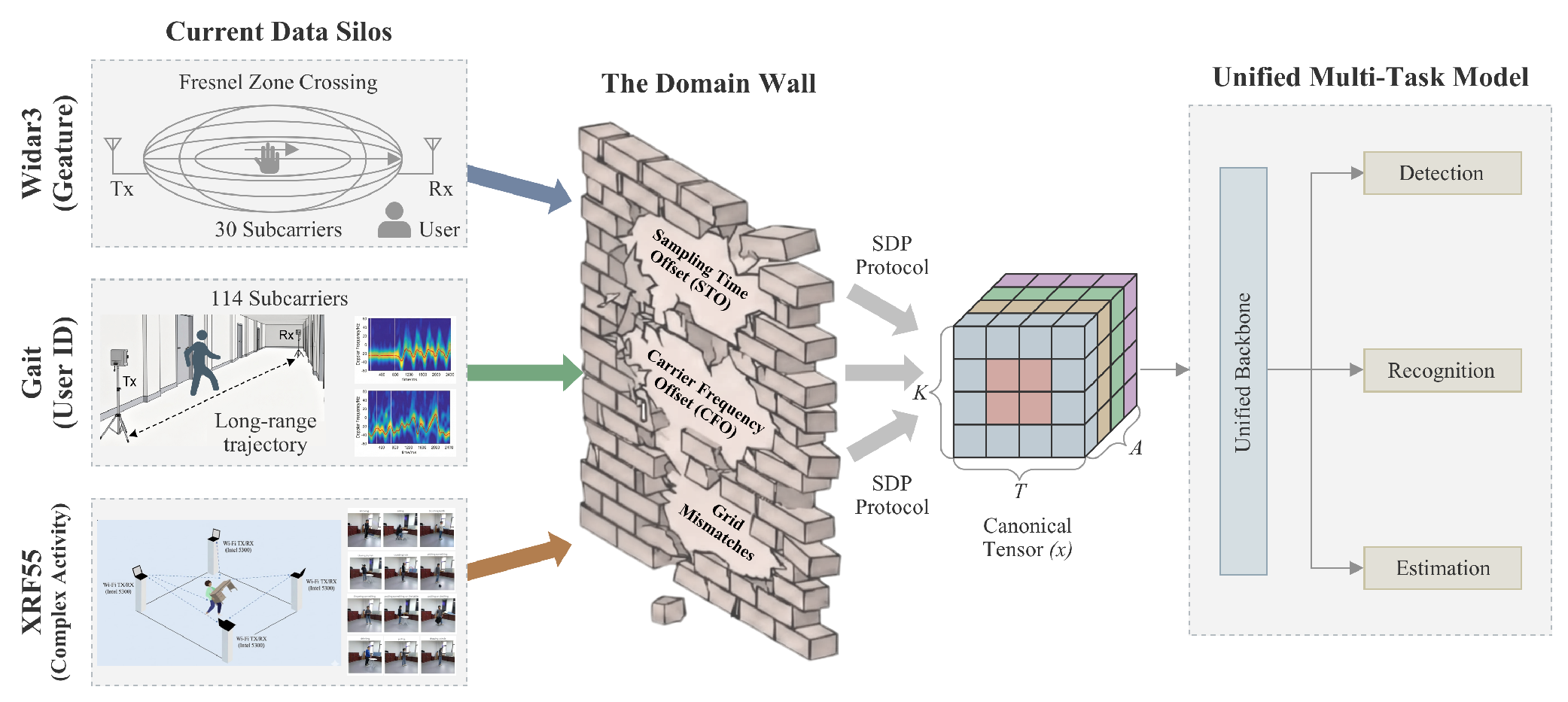}
    \caption{Protocol-level motivation of SDP. SDP canonicalizes heterogeneous CSI measurements from diverse devices/datasets into a standardized Canonical CSI Tensor, thereby reducing hardware-induced domain barriers and enabling unified, reproducible benchmarking across multiple sensing tasks under a common interface.}
    \label{fig:domain_wall}
    
\end{figure*}

    \item Building on SDP, we instantiate a multi-task, multi-dataset benchmark with a fixed Transformer-based backbone as a controlled probe (rather than a task-optimized model). Extensive evaluations across diverse sensing granularities (gesture, gait, and activity) demonstrate that SDP preserves competitive accuracy while substantially improving evaluation stability (e.g., reduced seed-level variance) and enabling fair comparisons across heterogeneous datasets and devices.
\end{itemize}

The remainder of this paper is organized as follows. Section~\ref{section2} reviews related datasets and research efforts on wireless sensing and ISAC benchmarking. Section~\ref{section3} presents the unified data abstraction and processing framework of SDP. Section~\ref{section4} introduces the benchmark architecture and experimental protocol. Section~\ref{section5} provides extensive experimental evaluations, and Section~\ref{section6} concludes the paper.

\section{Related Work}\label{section2}

\subsection{Wireless Sensing Datasets}

High-quality datasets constitute the foundation of data-driven ISAC systems and have significantly accelerated progress in wireless sensing, as summarized in Table~\ref{table-Related Datasets}. Early and widely adopted datasets, such as SignFi~\cite{ma2018signfi} and Widar3.0~\cite{9516988}, typically operate under specific hardware assumptions, with signal formats, annotations, and preprocessing pipelines tailored to commodity network interface cards (e.g., Intel 5300). These datasets have played a crucial role in validating the feasibility of CSI-based sensing under controlled settings. Subsequent efforts have expanded the sensing scope toward more complex scenarios, including multi-user interactions and multi-modal data fusion. For instance, EyeFi~\cite{9183685} introduced synchronized video as auxiliary ground truth to assist CSI trajectory analysis. Similarly, XRF V2~\cite{10.1145/3749521} has integrated Wi-Fi sensing with complementary modalities, including cameras, LiDAR, and wearable IMUs.

Despite the richness of these datasets, fundamental limitations persist. Most datasets are inherently device-centric, strictly tailored to specific hardware configurations (e.g., fixed sensor topologies or specific wearables). This results in heterogeneous subcarrier layouts, antenna configurations, sampling rates, packet formats, and metadata structures. Annotation schemes vary across tasks and datasets, preventing consistent evaluation. Preprocessing pipelines, including CSI sanitization, segmentation, and normalization, are implemented inconsistently, which undermines reproducibility and obscures the impact of modeling choices. Furthermore, datasets typically focus on a single sensing objective and do not provide a unified representation that enables fair comparison or cross-task learning. These persistent inconsistencies motivate the need for a canonical abstraction and standardized processing pipeline, which explicitly targets interoperability across devices, datasets, and sensing tasks.

\subsection{Preprocessing Pipelines}

Extracting robust features from raw CSI is critical for mitigating hardware impairments. Prior research has explored various sanitization pipelines. For example, UniFi~\cite{10.1145/3631429} highlighted the impact of amplitude noise and random phase offsets, proposing consistency-guided multi-view fusion to extract robust features. Other studies have employed phase calibration techniques, such as linear regression or conjugate multiplication, to remove random phase rotations caused by unsynchronized clocks.

However, the implementation of these preprocessing steps remains highly fragmented and heuristic. As noted in \cite{wang2025surveywifisensinggeneralizability}, there is no standardized protocol for critical operations like static component suppression or amplitude normalization. Existing studies often employ ad-hoc parameters or motion-triggered segmentation rules that are not reproducible across different hardware platforms. More critically, most preprocessing pipelines operate directly on raw heterogeneous IQ samples without normalizing them into a device-agnostic space. As a result, neural networks are forced to implicitly compensate for varying subcarrier layouts and hardware-specific artifacts, often leading to model overfitting and reduced robustness against domain shifts. This observation suggests that preprocessing should not be treated as an implementation detail, but rather as a first-class component of the learning pipeline that must be explicitly standardized.

\subsection{Generalizable Architectures and Multi-Task Evaluation}

A central challenge in learning-based wireless sensing lies in achieving generalizable and reproducible performance across heterogeneous sensing tasks, deployment environments, and hardware platforms.  Unlike vision-based perception, where standardized pixel representations largely decouple algorithms from sensors, wireless sensing models remain tightly coupled with device-specific CSI representations and preprocessing choices. As a result, performance improvements reported by learning-based approaches are often difficult to attribute solely to architectural advances, and their generalization across tasks and datasets remains unclear.

Driven by the rapid development of deep learning, a wide range of sensing architectures have been proposed to extract discriminative features from CSI streams. Early studies leveraged CNNs and RNNs (e.g., BiLSTM~\cite{8514811}, CNN-GRU~\cite{fu2024wi}) to capture local spatial patterns and temporal dependencies. More recent works adopted attention mechanisms and Transformer-based models to model global correlations across subcarriers and time steps. For example, WiTransformer~\cite{10477406} achieved 86.2\% accuracy on a 22-class gesture recognition benchmark, outperforming CNN--BiLSTM baselines by 5--10\%. State-space models have also been introduced to enable efficient long-sequence modeling, as demonstrated in XRF V2~\cite{10.1145/3749521}. In parallel, representation learning strategies such as adversarial learning and contrastive pre-training (e.g., WiFiGPT~\cite{bhatia2025transforming}, WirelessGPT~\cite{yang2025wirelessgpt}) have been explored to mitigate domain shifts and improve environment invariance.

Beyond individual model designs, recent efforts have begun to emphasize systematic evaluation and benchmarking. SenseFi~\cite{yang2023benchmark}, for instance, established a comprehensive benchmarking framework and software library for deep learning--based Wi-Fi sensing, providing a unified interface to evaluate diverse model families across multiple public datasets. By consolidating model implementations and training pipelines, SenseFi has facilitated algorithm-level comparison and accelerated empirical exploration.

Despite these advances, existing evaluation practices remain largely model-centric. Frameworks such as SenseFi primarily benchmark learning architectures under given CSI representations, typically relying on raw or lightly preprocessed channel measurements. Consequently, key sources of irreproducibility in practical wireless sensing systems, including hardware heterogeneity, vendor-specific CSI formats, and long-term signal instability, are implicitly absorbed by the learning model rather than being explicitly normalized or controlled. As a result, reported performance gains are often confounded by inconsistent preprocessing and evaluation settings, making it difficult to disentangle true algorithmic improvements from implementation-specific effects.

To address these limitations, SDP introduces a protocol-level abstraction for wireless sensing. Rather than emphasizing architectural exploration, SDP standardizes the entire sensing pipeline through a unified preprocessing middleware and a canonical tensor representation that explicitly normalizes hardware-dependent artifacts. By decoupling sensing tasks from device-specific CSI formats, SDP enables a unified learning protocol and a standardized multi-task benchmark. As a result, while existing frameworks primarily benchmark models under fixed input representations, SDP benchmarks and normalizes the sensing pipeline itself, enabling fair, reproducible, and cross-task evaluation across heterogeneous datasets, devices, and sensing scenarios.

\begin{figure*}[htp!]
    \centering
    \captionsetup{justification=raggedright}    
    \includegraphics[width=0.95\linewidth]{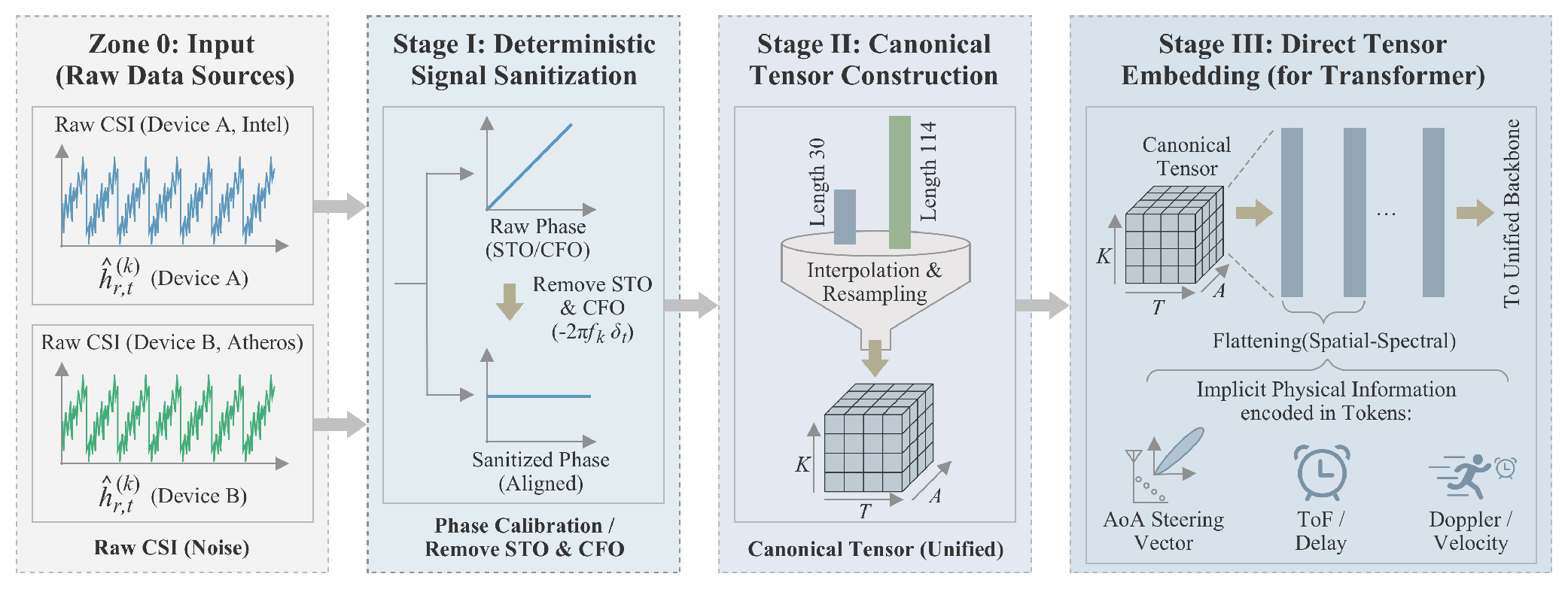}
    \caption{The SDP unified data processing pipeline. This figure details the SDP's deterministic workflow: Stage I removes hardware impairments like STO and CFO; Stage II projects diverse CSI into a uniform Canonical Tensor; Stage III flattens the tensor into a token sequence, ready for the Transformer backbone.}
    \label{fig:pipeline}
\end{figure*}

\section{System Model and Unified Protocol}\label{section3}

This section establishes the theoretical foundation of the SDP.
We first model the dominant hardware-induced impairments in raw wireless signals. We then present the SDP unified protocol, which defines a deterministic and reproducible procedure to transform heterogeneous CSI measurements into a standardized, learning-ready representation through signal sanitization and canonical tensor construction.

\subsection{Physical Signal Modeling and Hardware Impairments}

We consider an Orthogonal Frequency-Division Multiplexing (OFDM) system with $N_t$ transmit antennas and $N_r$ receive antennas\cite{liu2025sensingcommunicationsignalsinformation}.
As established in prior studies~\cite{yang2022hands}, the measured CSI $\hat{h}_{r,t}^{(k)}(t)$ at subcarrier $k$ deviates from the ideal physical channel due to unavoidable hardware imperfections.
The received signal model is expressed as
\begin{equation}
\begin{split}
\hat{h}_{r,t}^{(k)}(t) =
& \underbrace{e^{-j2\pi(f_k \delta_t + \epsilon_f t + \beta)}}_{\text{Hardware-Induced Phase Distortions}} \\
& \cdot \underbrace{\sum_{l=1}^{L} \alpha_l(t) e^{-j2\pi f_k \tau_l(t)} e^{j2\pi \nu_l(t) t}}_{\text{Physical Multipath Channel}}
+ n(t),
\end{split}
\label{eq:csi_model}
\end{equation}
where $\delta_t$ denotes the STO and packet detection delay (PDD), $\epsilon_f$ represents the CFO, and $\beta$ is the initial phase offset introduced by the phase-locked loop (PLL).
The second term captures the intrinsic physical propagation process, encoding geometric information through path delays $\tau_l$ and kinematic information through Doppler shifts $\nu_l$.

This formulation highlights the fundamental origin of the domain gap in wireless sensing.
Even when two devices observe the same physical event, their measured CSI streams can exhibit substantially different statistical characteristics due to device-dependent distortion parameters $(\delta_t, \epsilon_f, \beta)$ and heterogeneous sampling grids (e.g., different center frequencies and subcarrier spacings).
As a result, raw CSI measurements are not directly comparable across devices, rendering cross-domain learning unreliable without an explicit normalization protocol.

\subsection{Unified Protocol and Canonical Tensor Construction}

To standardize heterogeneous CSI measurements for downstream learning and evaluation, SDP defines a protocol-driven mapping procedure that transforms raw, hardware-dependent data into a device-agnostic canonical representation. 

\subsubsection{Deterministic Signal Sanitization}

The protocol first mitigates linear phase distortions associated with STO and PDD, as characterized in Eq.~(\ref{eq:csi_model}).
Specifically, a linear fitting procedure is applied to estimate the frequency-dependent phase slope $2\pi \delta_t f_k$ across subcarriers.
This estimated component is then removed from the measured CSI, aligning phase responses from different hardware platforms to a common reference.
As a result, subsequent phase variations are primarily attributable to physical channel dynamics rather than clock asynchrony or implementation-specific offsets.

\subsubsection{Canonical Frequency Projection}

A key source of heterogeneity in wireless sensing arises from mismatched subcarrier configurations across devices.
Let the raw CSI vector at a given time instant be denoted as $\mathbf{h}_{\text{raw}} \in \mathbb{C}^{K_{\text{raw}}}$.
SDP defines a canonical frequency resolution $K$ and projects $\mathbf{h}_{\text{raw}}$ onto a unified subcarrier grid via an interpolation operator
\begin{equation}
\mathbf{h}_{\text{canon}} = \mathcal{F}(\mathbf{h}_{\text{raw}}, K_{\text{raw}} \rightarrow K),
\label{eq:interpolation}
\end{equation}
thereby normalizing the spectral dimension regardless of the underlying hardware bandwidth or subcarrier density.

\subsubsection{Canonical Tensor Construction}

Following frequency projection, the CSI stream is segmented into fixed-length temporal windows of size $T$.
The spatial dimensions associated with transmit--receive antenna pairs are aggregated into a unified spatial mode of dimension $A = N_t \times N_r$.
The resulting samples are organized into the Canonical CSI Tensor
\begin{equation}
\mathcal{X} \in \mathbb{C}^{A \times K \times T}.
\label{eq:tensor}
\end{equation}

\subsubsection{Choice of Canonical Frequency Resolution.}
The canonical frequency grid size $k$ is a protocol-level design choice rather than a task-specific hyperparameter. It does not correspond to the native subcarrier count of any particular hardware platform, but instead defines a unified frequency resolution to enable fair and reproducible benchmarking across heterogeneous datasets and devices.

In this work, we fix the canonical grid size to $k=30$. This choice approximately matches the effective subcarrier resolution of widely used legacy CSI datasets collected with Intel~5300 NICs, while remaining sufficiently expressive to preserve discriminative frequency-selective patterns in modern Wi-Fi~6 measurements with substantially higher raw subcarrier counts. By mapping raw CSI measurements with arbitrary bandwidths and FFT sizes onto a fixed-length canonical grid, SDP eliminates hardware-dependent frequency resolution as a confounding factor in learning-based sensing evaluation.

The selection of $k=30$ reflects a trade-off between representation sufficiency and benchmarking stability. A larger canonical grid (e.g., directly retaining the native subcarrier resolution with $k=512$) reintroduces hardware-specific noise, amplifies phase-related imperfections, and increases sensitivity to random initialization and optimization dynamics. Conversely, an overly coarse grid would excessively smooth frequency-domain structures critical for sensing tasks. As validated in the ablation study, increasing $k$ beyond the canonical resolution may yield sporadic gains in single-run accuracy, but leads to significantly higher inter-seed variance and degraded reproducibility.

Therefore, $k=30$ represents a minimal yet sufficient canonical frequency resolution that balances expressiveness and stability, aligning with SDP’s objective of establishing a reliable benchmarking middleware rather than optimizing for task-specific peak performance.

The canonical tensor $\mathcal{X}$ constitutes the fundamental representation defined by the SDP protocol. All benchmark models and evaluation procedures operate directly on this representation, which is device-agnostic by design. This design explicitly preserves the complete signal structure, enabling downstream models to learn robust features end-to-end without information loss, while ensuring that protocol compliance remains independent of device-specific configurations.

\begin{table*}[htp!]
\centering
\caption{Overview of Datasets in our SDP benchmark}
\label{tab:datasets}
\begin{tblr}{
  width = \linewidth,
  colspec = {Q[102]Q[140]Q[142]Q[96]Q[240]Q[40]Q[56]Q[100]},
  row{1} = {c},
  cell{2}{1} = {c},
  cell{2}{2} = {c},
  cell{2}{3} = {c},
  cell{2}{4} = {c},
  cell{2}{6} = {c},
  cell{2}{7} = {c},
  cell{2}{8} = {c},
  cell{3}{1} = {c},
  cell{3}{2} = {c},
  cell{3}{3} = {c},
  cell{3}{4} = {c},
  cell{3}{6} = {c},
  cell{3}{7} = {c},
  cell{3}{8} = {c},
  cell{4}{1} = {c},
  cell{4}{2} = {c},
  cell{4}{3} = {c},
  cell{4}{4} = {c},
  cell{4}{6} = {c},
  cell{4}{7} = {c},
  cell{4}{8} = {c},
  cell{5}{1} = {c},
  cell{5}{2} = {c},
  cell{5}{3} = {c},
  cell{5}{4} = {c},
  cell{5}{6} = {c},
  cell{5}{7} = {c},
  cell{5}{8} = {c},
  hlines,
  vlines,
}
\textbf{Dataset}     & \textbf{Sensing Task}                              & \textbf{Motion Complexity}                       & \textbf{Hardware}    & \textbf{Collection Scenario}                                                                         & \textbf{\#Subj.} & \textbf{\#Act.} & \textbf{Total Samples} \\
\textbf{Widar3.0} \cite{9516988}    & Gesture Recognition                                & Short, fine-grained motions                      & Intel 5300           & Multiple indoor environments including classrooms, corridors, and office spaces with varying layouts & 16               & 9–15            & $\sim$260,000          \\
\textbf{GaitID}\cite{meng2020gait}      & Gait-based User Identification                     & Periodic micro-motions                           & Intel 5300           & Controlled indoor walking trajectories in corridors and meeting rooms                                & 11               & {1\\(walking)}  & $\sim$2,000            \\
\textbf{XRF55} \cite{10.1145/3643543}       & Fine-grained Human Activity Recognition            & Long-duration, compositional activities          & Intel 5300           & Furnished indoor environments (laboratories, living rooms, corridors) with complex object layouts    & 39               & 55              & 134,476                \\
\textbf{ElderAL-CSI} \cite{DBLP:data/11/LiHCSC25} & Joint Activity Recognition and Indoor Localization & Confusable daily activities and emergency events & ZTE AX3000 (Wi-Fi~6) & Large-scale open indoor environment (172~m$^2$), unfurnished, with grid-based spatial deployment     & 3                & 6               & 42,147                 
\end{tblr}
\end{table*}

\section{Benchmark Backbone}\label{section4}

Building upon the unified protocol established in Section~\ref{section3}, this section defines the unified benchmark architecture. The objective is not to propose a complex, task-specific novelty, but to establish a rigorous, reproducible, and extensible evaluation baseline. This framework operates directly on the Canonical CSI Tensor, utilizing a consistent neural backbone to enable fair comparison across diverse hardware platforms and sensing granularities without heuristic manual tuning.

\subsection{Benchmark Formulation and Input Interface}

Let $\mathcal{X} \in \mathbb{C}^{A \times K \times T}$ denote the Canonical CSI Tensor produced by the standardized protocol. Unlike prior works that rely on handcrafted feature extraction (e.g., statistical pooling or tensor decomposition), our benchmark adopts a direct tensor embedding strategy.

We treat the CSI tensor as a sequence of tokens, utilizing an architecture analogous to Vision Transformers (ViT). In this analogy, each temporal snapshot of the CSI stream serves as a token, encapsulating the instantaneous spatial-spectral state, similar to how an image patch captures local visual texture. Specifically, the tensor is flattened along the spatial and spectral dimensions at each time step, resulting in a sequence $\mathbf{X}_{seq} \in \mathbb{R}^{T \times (A \cdot K)}$. A linear projection layer maps this sequence into a latent embedding space of dimension $D$:
\begin{equation}
    \mathbf{E} = \mathbf{X}_{seq} \mathbf{W}_{proj} + \mathbf{E}_{pos},
\end{equation}
where $\mathbf{W}_{proj} \in \mathbb{R}^{(A \cdot K) \times D}$ is a learnable projection matrix, and $\mathbf{E}_{pos} \in \mathbb{R}^{T \times D}$ denotes learnable positional encodings that preserve temporal order information. This design ensures that the input retains the complete physical semantics (phase and amplitude variations over time) while adapting to a standard deep learning format.

\subsection{Unified Reference Architecture}

To provide a transparent baseline, SDP instantiates the framework with a streamlined unified transformer backbone. Crucially, we enforce a consistent architectural configuration across all sensing tasks.

\subsubsection{Unified Transformer Backbone}

The backbone is designed to be task-agnostic, prioritizing generalizability over complexity. It consists of $L$ stacked Transformer Encoder layers. Each layer comprises a Multi-Head Self-Attention (MSA) block and a Feed-Forward Network (FFN), incorporating residual connections and Layer Normalization (LN) as follows:
\begin{equation}
\begin{aligned}
    \mathbf{Z}' &= \text{MSA}(\text{LN}(\mathbf{Z}_{l-1})) + \mathbf{Z}_{l-1}, \\
    \mathbf{Z}_l &= \text{FFN}(\text{LN}(\mathbf{Z}')) + \mathbf{Z}'.
\end{aligned}
\end{equation}
This architecture enables the model to capture global temporal correlations within the CSI stream. To ensure benchmark reproducibility, the hyperparameters (e.g., layers $L$, embedding dimension $D$, heads $H$) are locked to a fixed standard configuration. A global average pooling operation is finally applied to the output sequence to produce the task-agnostic latent representation $\mathbf{z} \in \mathbb{R}^D$.

\subsubsection{Task-Specific Projection Heads}
While the backbone remains frozen in structure, the final output head is adapted to the specific sensing objective via a minimal Linear Projection Head $g_\phi(\cdot)$. For discrete classification tasks such as gesture recognition and gait identification, we employ a linear transformation followed by a Softmax activation, formulated as $\mathbf{y} = \text{Softmax}(\mathbf{W}_{cls} \mathbf{z} + \mathbf{b})$. In contrast, for continuous regression objectives like trajectory tracking or activity estimation, a direct linear mapping $\mathbf{y} = \mathbf{W}_{reg} \mathbf{z} + \mathbf{b}$ is utilized. This distinction is crucial as it demonstrates that the unified latent representation $\mathbf{z}$ is semantically rich enough to support both categorical decision boundaries and continuous metric estimation without backbone modification. This design allows us to evaluate the generality of the canonical representation: if the same backbone structure achieves robust and consistent performance across radically different tasks, it validates the efficacy of the SDP protocol.

\begin{algorithm}[htp!]
\caption{Unified Benchmark Training Procedure}
\label{alg:sdp_procedure}
\begin{algorithmic}[1]
\REQUIRE Standardized dataset $\mathcal{D}_{\text{std}} = \{(\mathcal{X}_i, y_i)\}_{i=1}^N$ for a specific task $\tau$; \\
Backbone config: Layers $L$, Dim $D$, Heads $H$; \\
Training config: Epochs $E$, Batch size $B$, Learning rate $\eta$.
\ENSURE Trained model $f_\theta$ and head $g_\phi$.

\STATE \textbf{Initialization:}
\STATE Construct backbone $f_\theta$ with config $(L, D, H)$.
\STATE Attach task head $g_\phi$ (Classification or Regression).
\STATE Set random seed $\mathcal{S}_{seed}$ for reproducibility.

\STATE \textbf{Training Loop:}
\FOR{epoch $e = 1$ to $E$}
    \STATE Shuffle $\mathcal{D}_{\text{std}}$.
    \FOR{each mini-batch $(\mathbf{X}_{batch}, \mathbf{y}_{batch}) \in \mathcal{D}_{\text{std}}$}
        \STATE $\mathbf{E} \leftarrow \textsc{PatchEmbed}(\mathbf{X}_{batch})$ \COMMENT{Flatten \& Project}
        \STATE $\mathbf{z} \leftarrow f_\theta(\mathbf{E})$ \COMMENT{Unified Backbone Forward}
        \STATE $\hat{\mathbf{y}} \leftarrow g_\phi(\mathbf{z})$ \COMMENT{Task Head Prediction}
        
        \IF{task is Classification}
            \STATE $\mathcal{L} \leftarrow \text{CrossEntropy}(\hat{\mathbf{y}}, \mathbf{y}_{batch})$
        \ELSE
            \STATE $\mathcal{L} \leftarrow \text{MSE}(\hat{\mathbf{y}}, \mathbf{y}_{batch})$
        \ENDIF
        
        \STATE Update $\theta, \phi$ via AdamW$(\nabla \mathcal{L}, \eta_e)$
    \ENDFOR
    \STATE Update $\eta_e$ via Cosine Scheduler.
\ENDFOR

\RETURN Best model $\{f_\theta, g_\phi\}$ based on validation set.
\end{algorithmic}
\end{algorithm}

\subsection{Standardized Training Protocol}

To eliminate implementation-dependent variance, a major source of unreproducibility in wireless sensing, SDP enforces a standardized independent training strategy. Rather than jointly optimizing multiple tasks, which often introduces complex loss balancing issues, each model is trained from scratch for a specific task $\tau \in \mathcal{T}$ following the deterministic procedure summarized in Algorithm~\ref{alg:sdp_procedure}. Crucially, we standardize all critical configuration factors to ensure a controlled evaluation environment: the AdamW optimizer with fixed weight decay is utilized in conjunction with a Cosine Annealing learning rate schedule; data augmentation is explicitly disabled to strictly isolate the impact of data quality; and fixed random seeds are enforced for weight initialization. By locking these "moving parts," SDP ensures that any observed performance differences are primarily attributable to the efficacy of the data protocol and the intrinsic difficulty of the task, rather than stochastic tuning tricks.

\section{Experiments and Results}\label{section5}

\subsection{Experimental Setup}

All experiments are conducted under a strictly unified benchmark configuration to evaluate the proposed SDP protocol in a fair, reproducible, and protocol-centric manner. As summarized in Table~\ref{tab:datasets}, the selected datasets intentionally span diverse sensing tasks, motion complexities, collection scenarios, and hardware generations. Rather than optimizing performance for any individual task or dataset, the experimental setup is designed to assess whether a single standardized sensing pipeline can support consistent and comparable evaluation across heterogeneous wireless sensing problems.

To prevent inter-session and inter-user information leakage, a strict cross-user split strategy is adopted for all datasets. Evaluation is primarily performed at the window level following the canonical segmentation defined by SDP, with clip-level aggregation reported only for reference when applicable. This ensures that reported results reflect subject-level generalization rather than memorization of user-specific patterns.

\begin{figure*}[t!] 
    \centering
    \captionsetup{justification=raggedright}

    \begin{subfigure}[b]{0.52\textwidth}  
        \centering
        \includegraphics[width=\linewidth]{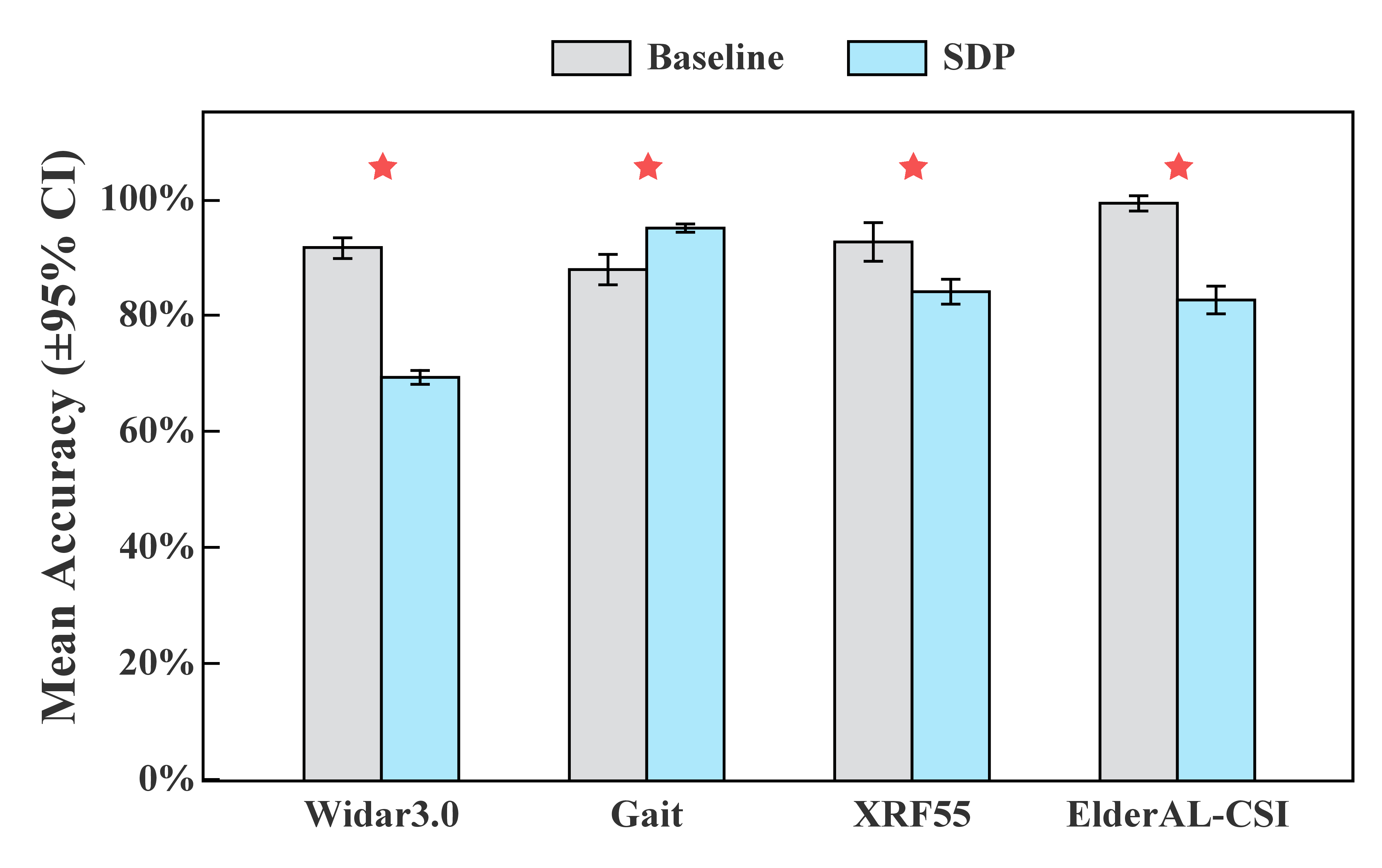} 
        \caption{Mean Accuracy Comparison}
        \label{fig:acc}
    \end{subfigure}
    \hfill
    \begin{subfigure}[b]{0.41\textwidth} 
        \centering
        \begin{subfigure}[b]{0.48\linewidth}
            \includegraphics[width=\linewidth]{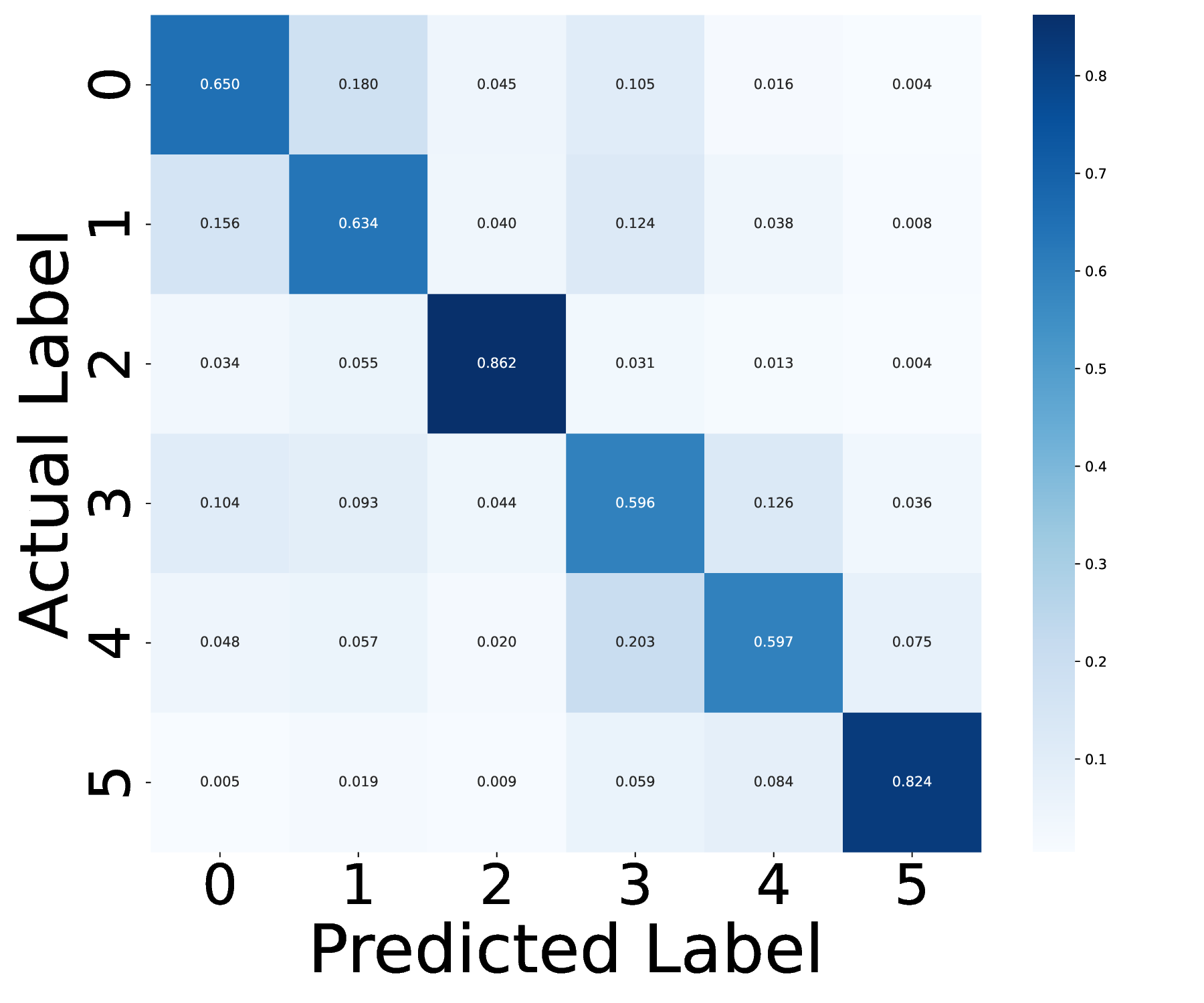}
            \caption{Widar3.0}
            \label{fig:cm_widar}
        \end{subfigure}
        \hfill
        \begin{subfigure}[b]{0.48\linewidth}
            \includegraphics[width=\linewidth]{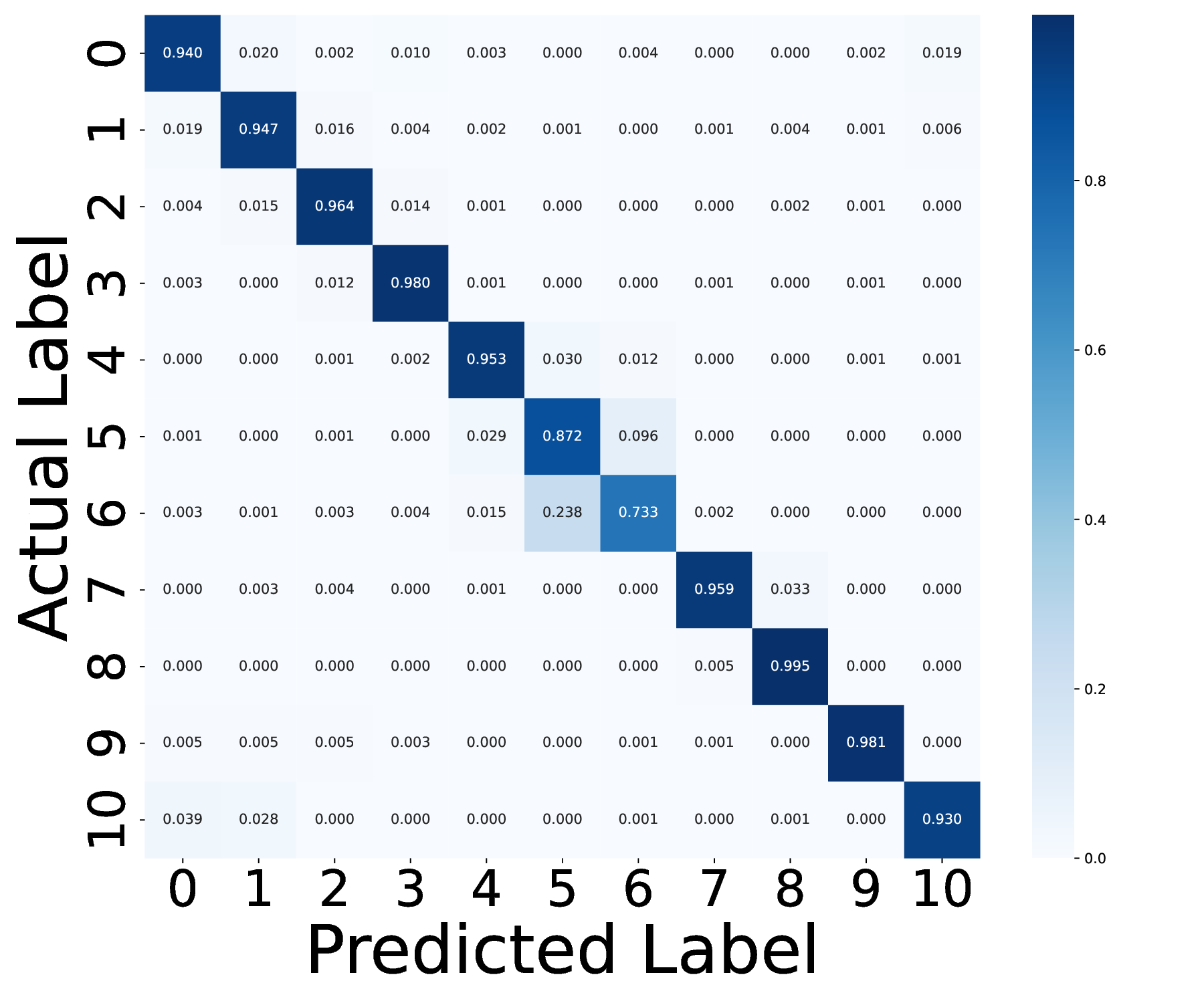}
            \caption{GaitID}
            \label{fig:cm_gait}
        \end{subfigure}
        
        \vspace{1pt} 
        
        \begin{subfigure}[b]{0.48\linewidth}
            \includegraphics[width=\linewidth]{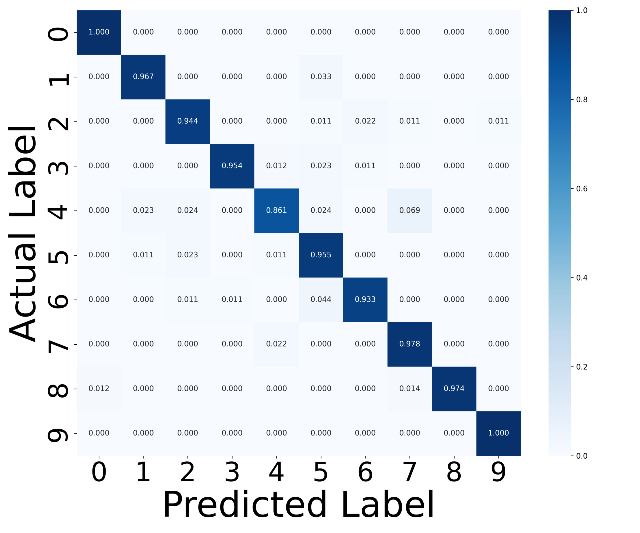}
            \caption{XRF55}
            \label{fig:cm_xrf55}
        \end{subfigure}
        \hfill
        \begin{subfigure}[b]{0.48\linewidth}
            \includegraphics[width=\linewidth]{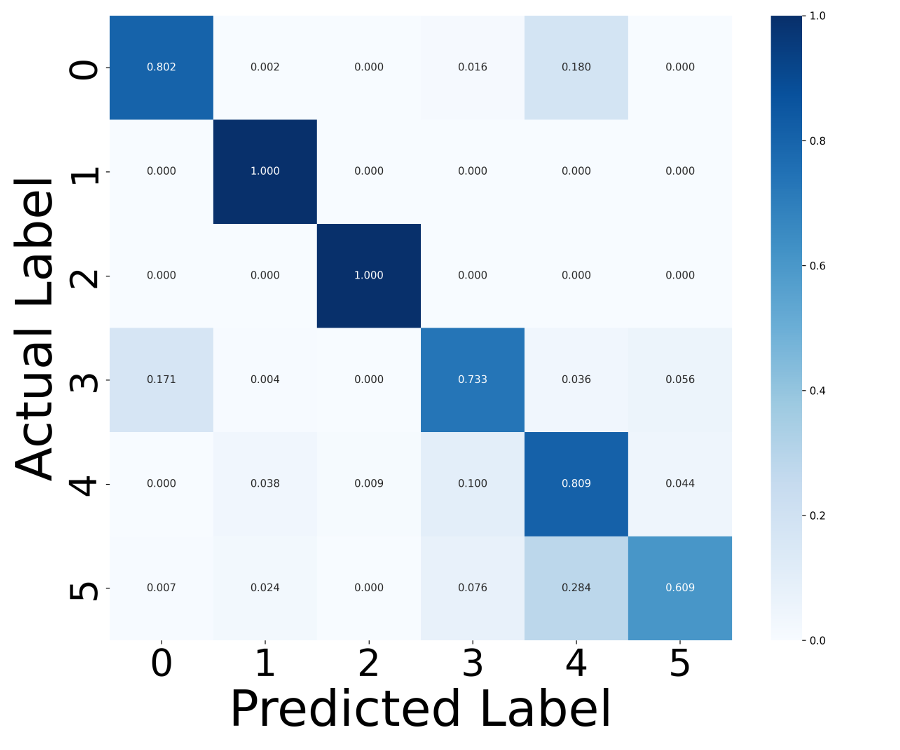}
            \caption{ElderAL-CSI}
            \label{fig:cm_elderal}
        \end{subfigure}
        \label{fig:heatmap}
    \end{subfigure}

    \vspace{1pt} 
    \caption{Performance of the SDP benchmark across heterogeneous sensing tasks. (a) Mean Top-1 accuracy with 95\% confidence intervals over five runs. (b--e) Normalized confusion matrices on Widar3.0, GaitID, XRF55, and ElderAL-CSI, showing consistent diagonal dominance across datasets.}
    \label{fig:Main Results}
\end{figure*}

Following the limited-tuning principle, a single unified model configuration and training protocol are enforced across all datasets and tasks. All inputs are mapped to the canonical CSI tensor defined by SDP and processed using an identical Transformer-based backbone with fixed architectural parameters (e.g., depth $L$ and embedding dimension $D$). The backbone is deliberately treated as a fixed probe rather than a performance-optimized model, and no dataset-specific architectural engineering or task-dependent hyperparameter tuning is introduced. Model optimization is performed using AdamW with a cosine annealing learning-rate schedule and early stopping based on validation performance.

To quantify statistical robustness and reproducibility, each experiment is repeated using five fixed random seeds $\mathcal{S}=\{992, 863, 702, 443, 542\}$. Performance metrics are reported as mean $\pm$ standard deviation across independent runs, with 95\% confidence intervals computed via the Student-$t$ distribution. Top-1 accuracy and macro-F1 score are used for classification tasks (e.g., gesture recognition and user identification), while Mean Absolute Error (MAE) is reported for regression-oriented tasks (e.g., tracking or localization).

All experiments are conducted on a single NVIDIA A100 GPU (40~GB) using CUDA~12.0 and PyTorch~2.0. Computational efficiency is evaluated in terms of parameter count, FLOPs, peak GPU memory usage, and single-window inference latency (p50/p90, batch size $=1$, post-warm-up). The choice of computing hardware is orthogonal to the SDP protocol and is reported solely for reproducibility.

\begin{figure}[hbp!]
    \centering
    \captionsetup{justification=raggedright}   
    \includegraphics[width=1\linewidth]{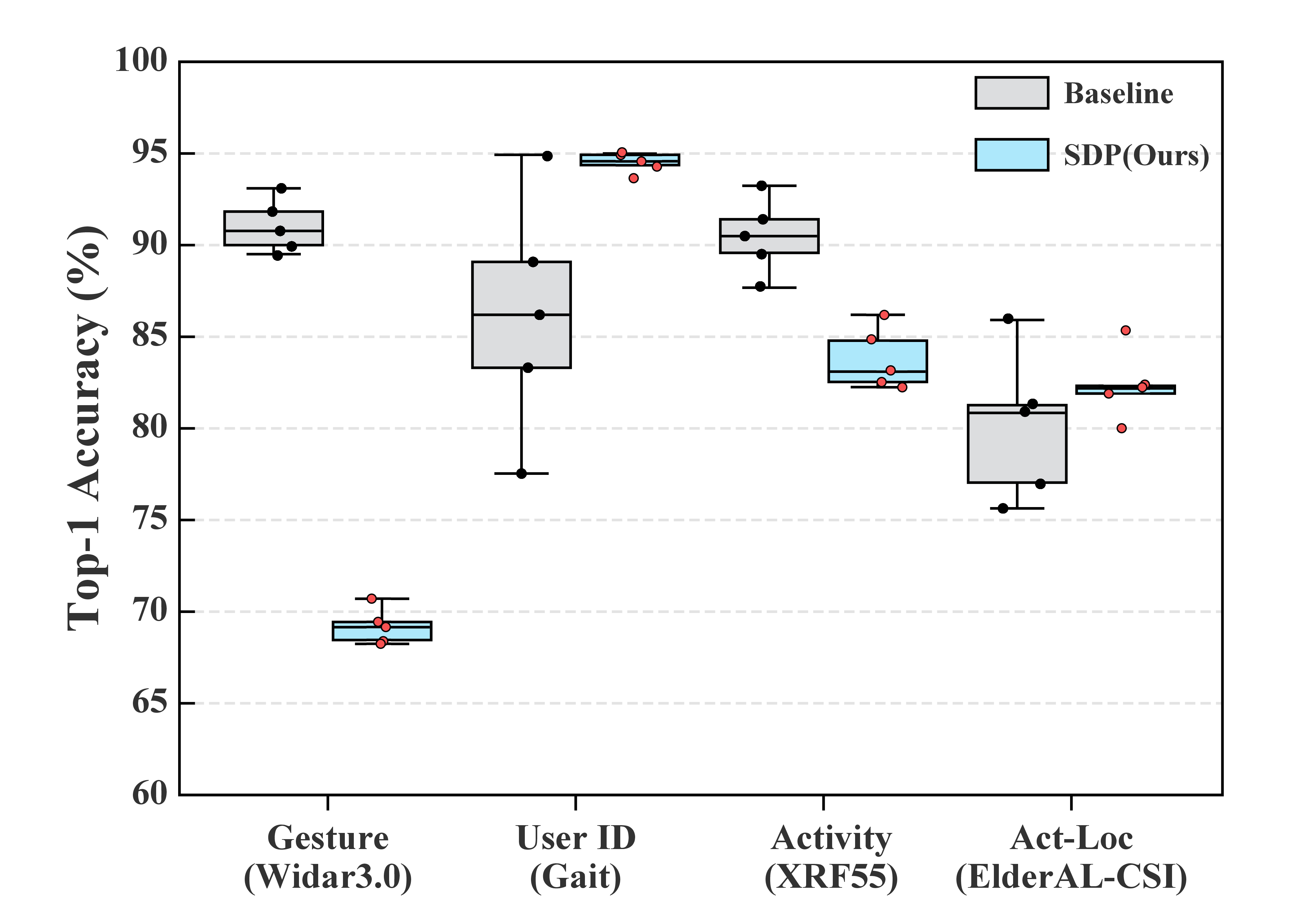}
    \caption{Performance stability comparison between the baseline and SDP across five random seeds. Boxplots show the distribution of Top-1 accuracy, with scattered dots indicating individual runs.}
    \label{fig:stability}
\end{figure}

\begin{figure*}[hbp!]
    \centering
    \captionsetup{justification=raggedright}

    \begin{subfigure}[b]{0.32\textwidth}
        \centering
        \includegraphics[width=0.49\linewidth]{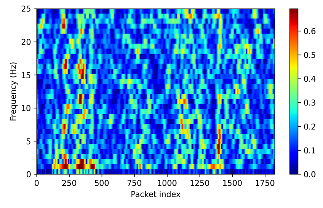} \hfill \includegraphics[width=0.49\linewidth]{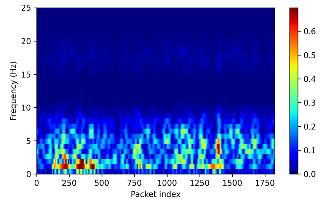}
        \caption{Sweep (Widar3.0)} 
        \label{fig:widar_sweep}
    \end{subfigure}
    \hfill
    \begin{subfigure}[b]{0.32\textwidth}
        \centering
        \includegraphics[width=0.49\linewidth]{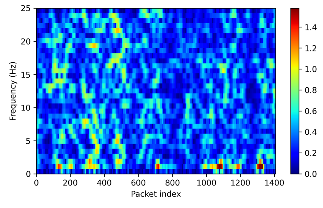} \hfill \includegraphics[width=0.49\linewidth]{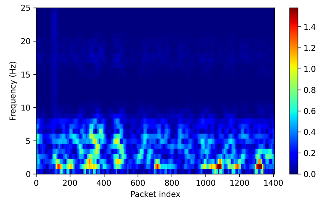}
        \caption{Clap (Widar3.0)}
        \label{fig:widar_clap}
    \end{subfigure}
    \hfill
    \begin{subfigure}[b]{0.32\textwidth}
        \centering
        \includegraphics[width=0.49\linewidth]{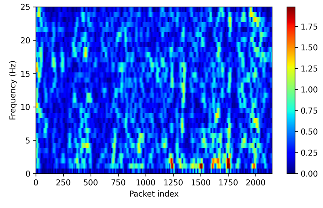} \hfill \includegraphics[width=0.49\linewidth]{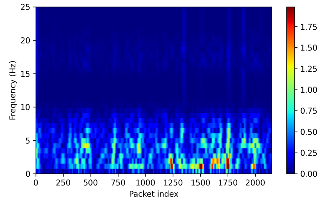}
        \caption{Draw Circle (Widar3.0)}
        \label{fig:widar_circle}
    \end{subfigure}

    \vspace{1pt} 

    \begin{subfigure}[b]{0.32\textwidth}
        \centering
        \includegraphics[width=0.49\linewidth]{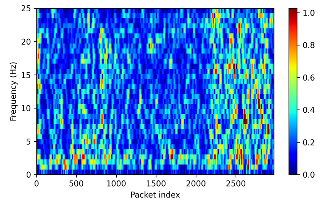} \hfill \includegraphics[width=0.49\linewidth]{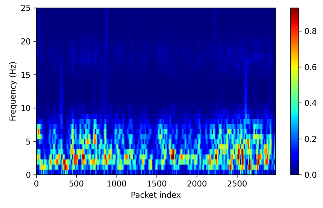}
        \caption{User A Walking (GaitID)}
        \label{fig:gait_userA}
    \end{subfigure}
    \hfill
    \begin{subfigure}[b]{0.32\textwidth}
        \centering
        \includegraphics[width=0.49\linewidth]{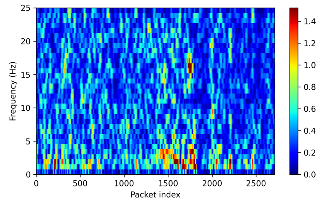} \hfill \includegraphics[width=0.49\linewidth]{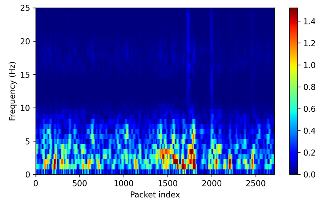}
        \caption{User B Walking (GaitID)}
        \label{fig:gait_userB}
    \end{subfigure}
    \hfill
    \begin{subfigure}[b]{0.32\textwidth}
        \centering
        \includegraphics[width=0.49\linewidth]{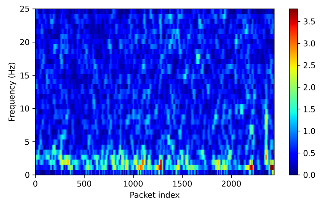} \hfill \includegraphics[width=0.49\linewidth]{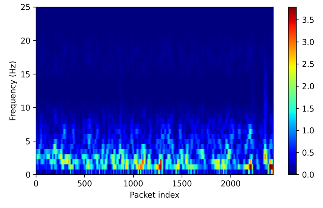}
        \caption{User C Walking (GaitID)}
        \label{fig:gait_userC}
    \end{subfigure}
    
    \vspace{1pt}

    \begin{subfigure}[b]{0.32\textwidth} 
        \centering
        \includegraphics[width=0.49\linewidth]{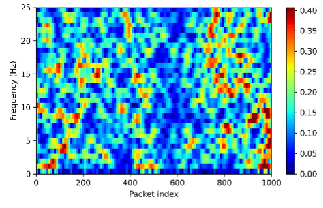} \hfill \includegraphics[width=0.49\linewidth]{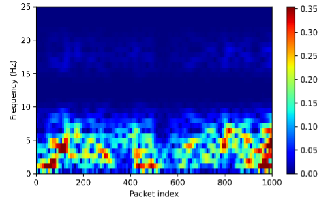}
        \caption{Picking (XRF55)}
        \label{fig:xrf_picking}
    \end{subfigure}
    \hfill
    \begin{subfigure}[b]{0.32\textwidth}
        \centering
        \includegraphics[width=0.49\linewidth]{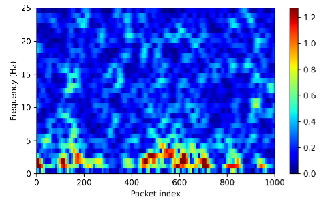} \hfill \includegraphics[width=0.49\linewidth]{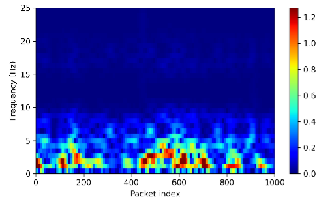}
        \caption{Hugging (XRF55)}
        \label{fig:xrf_hugging}
    \end{subfigure}
    \hfill
    \begin{subfigure}[b]{0.32\textwidth}
        \centering
        \includegraphics[width=0.49\linewidth]{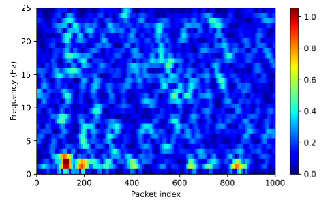} \hfill \includegraphics[width=0.49\linewidth]{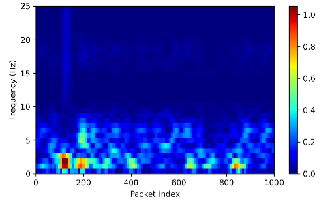}
        \caption{Running (XRF55)}
        \label{fig:xrf_running}
    \end{subfigure}
    \caption{Representation-level ablation via DFS spectrogram visualizations across heterogeneous tasks. Rows correspond to datasets (Widar3.0, GaitID, XRF55). In each subfigure, the left panel shows the raw spectrogram, while the right panel shows the SDP-processed representation. SDP consistently suppresses hardware-induced noise and enhances motion-related structures across tasks.}
    \label{fig:ablation_spectrograms}
\end{figure*}

This experimental setup is intentionally designed to minimize implementation-dependent variance, such that any observed performance differences can be primarily attributed to the standardized data protocol rather than task-specific model engineering. This directly aligns with the central objective of SDP as a unified preprocessing middleware and multi-task, multi-dataset benchmark, rather than a task-specific sensing model.

\subsection{Experimental Results}

\subsubsection{Main Results Across Tasks}

We evaluate SDP against task-specific native baselines across heterogeneous sensing tasks, including gesture recognition (Widar3.0), gait-based user identification (GaitID), complex activity recognition (XRF55), and joint activity--location recognition on ElderAL-CSI. The quantitative comparison is summarized in Fig.~\ref{fig:acc}. To further analyze model behavior beyond scalar metrics, Fig.~\ref{fig:cm_widar}--\ref{fig:cm_elderal} provide complementary visualizations of prediction errors from different perspectives.

Fig.~\ref{fig:acc} shows that under a single unified benchmark configuration, the impact of SDP on mean accuracy is inherently task-dependent rather than uniformly positive. This behavior is expected and consistent with SDP’s positioning as a benchmarking middleware that standardizes the sensing pipeline, rather than a task-specific optimizer tuned for peak performance. Across datasets, the most consistent trend is a systematic reduction in run-to-run uncertainty, reflected by tighter confidence intervals and more concentrated seed-wise distributions. Notably, on ElderAL-CSI, SDP remains effective under modern Wi-Fi~6 CSI characteristics and a joint activity--location label space, indicating that the unified preprocessing interface generalizes beyond legacy Intel~5300-style benchmarks.

Beyond mean performance, Fig.~\ref{fig:cm_widar}--\ref{fig:cm_elderal}  presents normalized confusion matrices that characterize the structural patterns of prediction errors under the unified SDP pipeline. For Widar3.0 and GaitID, the confusion patterns are sparse with probability mass concentrated along the diagonal, indicating stable class separation. For XRF55, which involves complex and highly confusable daily activities, off-diagonal errors become more pronounced, yet remain structured rather than random, suggesting that SDP preserves meaningful inter-class relationships under a standardized input interface. For clarity, the XRF55 visualization shows a representative 10-class subset spanning multiple activity groups, while all quantitative evaluations are performed on the full 55-class task. For ElderAL-CSI, stronger off-diagonal structures reflect the intrinsic difficulty of emergency-related activities coupled with location context.

Collectively, these results indicate that SDP provides a consistent and interpretable evaluation interface across diverse datasets, preserving task-dependent difficulty characteristics while enabling statistically reliable cross-task comparison under a unified benchmark.

\subsubsection{Reproducibility and Stability Analysis}

To assess SDP as a benchmarking middleware rather than a task-specific optimization technique, we investigate the stability and reproducibility of model performance under repeated training with different random seeds. Fig.~\ref{fig:stability} reports the distribution of Top-1 accuracy over five independent runs across four representative datasets, comparing task-specific native preprocessing pipelines with the unified SDP protocol.

\begin{figure*}[htp!]
    \centering
    \newlength{\commonheight}
    \setlength{\commonheight}{5cm} 
    \begin{subfigure}[t]{0.48\textwidth}
        \centering
        \includegraphics[width=\linewidth, height=\commonheight, keepaspectratio]{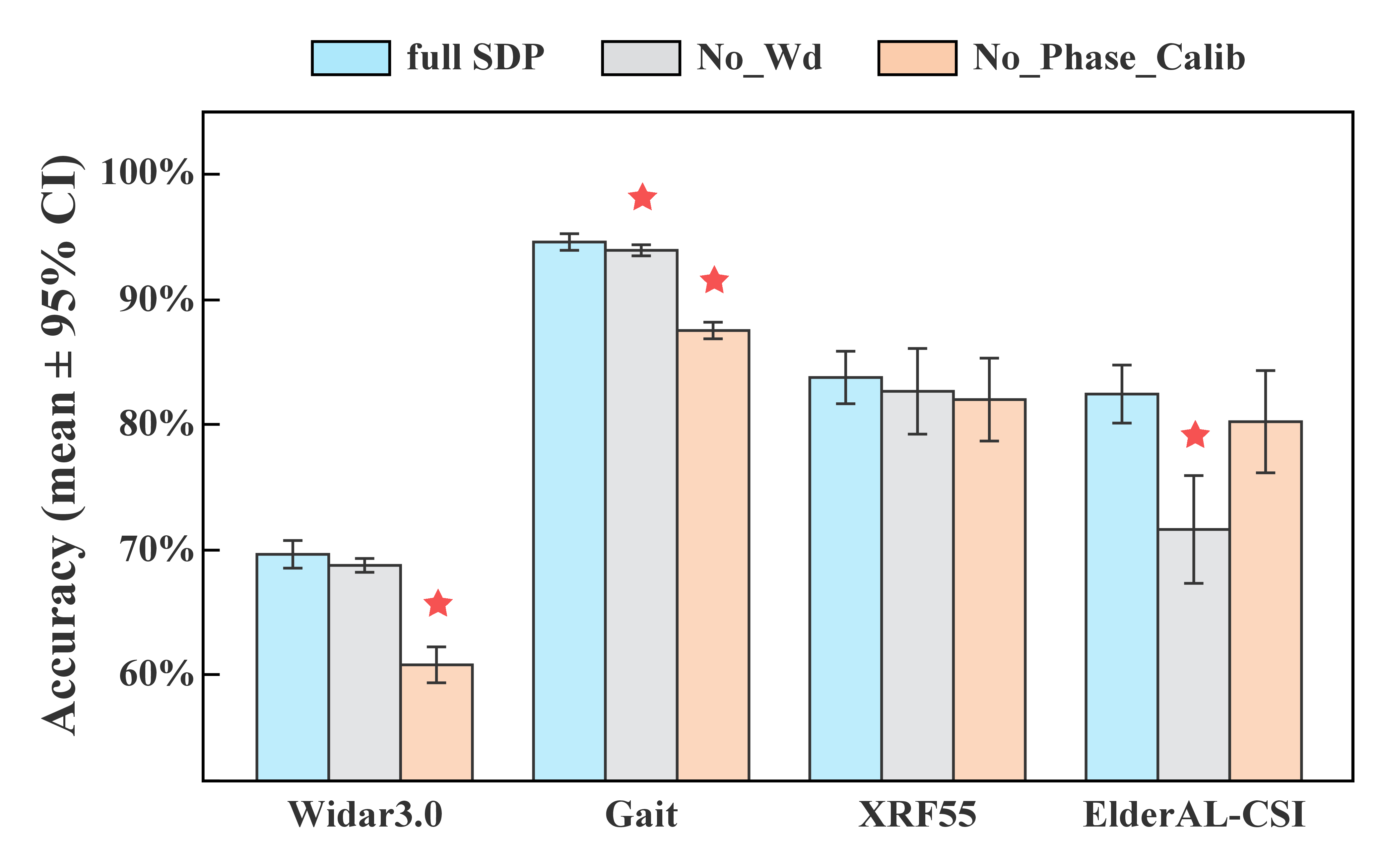}
        \captionsetup{justification=raggedright}
        \caption{Seed-level ablation on four datasets respectively. Individual bars show average Top-1 accuracy across five seeds, error bars denote the standard deviation and red dots indicate the 95\% confidence intervals. The full SDP pipeline exhibits stable performance, while removing key components results in increased variance.}
        \label{fig:ablation_seeds}
    \end{subfigure}
    \hfill 
    \begin{subfigure}[t]{0.48\textwidth}
        \centering
        \includegraphics[width=\linewidth, height=\commonheight, keepaspectratio]{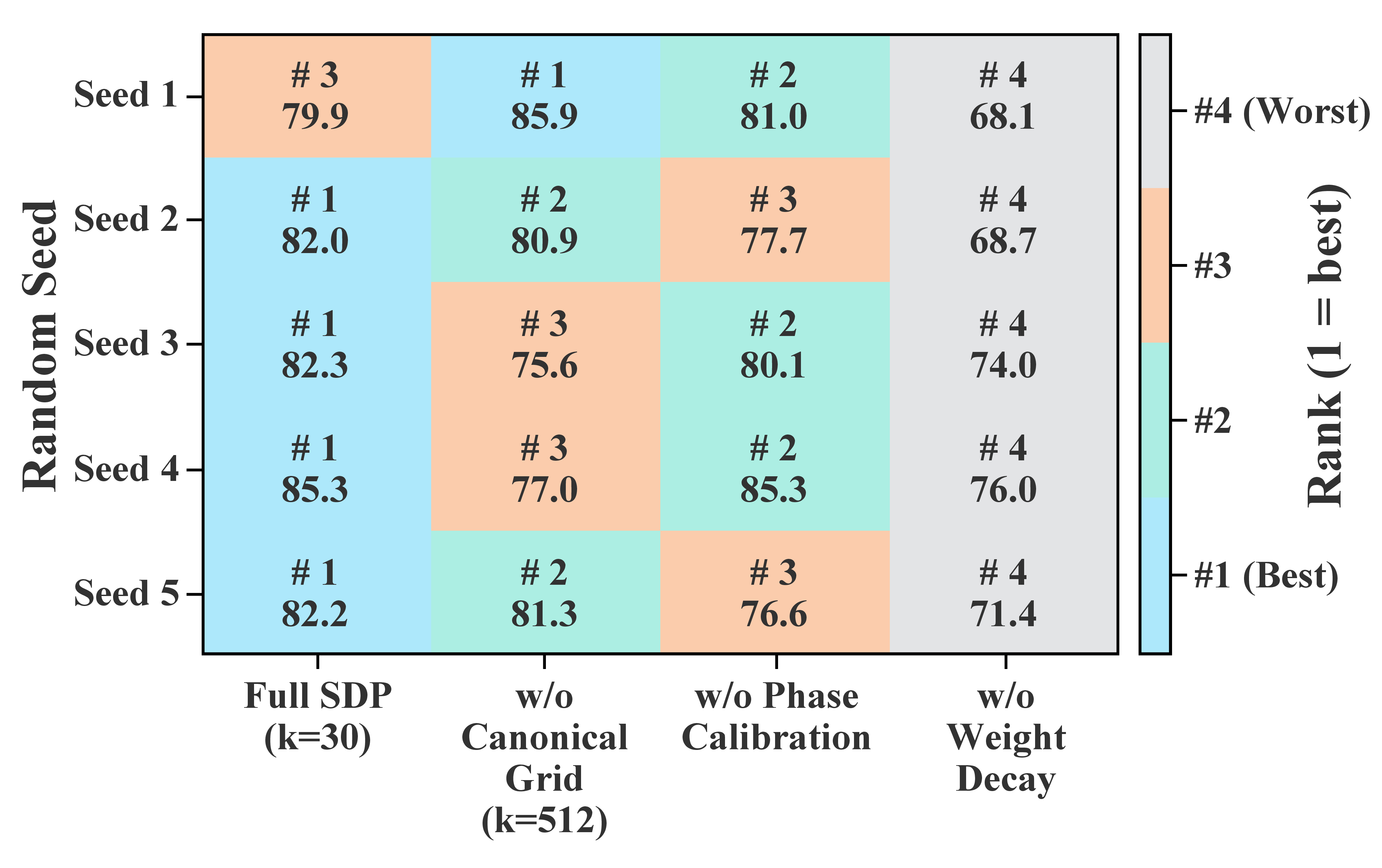}
        \captionsetup{justification=raggedright}
        \caption{Rank consistency heatmap across five random seeds on the ElderAL-CSI dataset. Colors indicate per-seed performance rank (1 = best), with overlaid Top-1 accuracy values. Full SDP exhibits stable top-ranked performance, while ablated variants show higher ranking variability.}
        \label{fig:ablation_rank}
    \end{subfigure}

    \caption{Ablation study results analyzing performance stability across datasets (a) and rank consistency on ElderAL-CSI (b).}
    \label{fig:combined_ablation_results}
\end{figure*}

Across all tasks, a consistent trend is observed: while mean accuracy varies with task characteristics, SDP substantially reduces inter-seed variance under identical training configurations. For relatively structured tasks such as gesture recognition on Widar3.0, the native baseline already achieves high accuracy, yet SDP further tightens the performance distribution, indicating reduced sensitivity to initialization and optimization randomness. This stabilization effect becomes more pronounced for tasks with higher intrinsic complexity. On GaitID and XRF55, native pipelines exhibit wide performance fluctuations across runs, whereas SDP yields compact, well-concentrated distributions, corresponding to a variance reduction of more than one order of magnitude in some cases.

The effect is particularly evident on ElderAL-CSI, which involves joint activity--location recognition under modern Wi-Fi~6 hardware. Here, the native pipeline shows strong sensitivity to both hardware characteristics and label ambiguity, resulting in unstable outcomes across seeds. In contrast, SDP maintains a concentrated performance distribution despite operating under the same unified configuration, demonstrating that its stabilizing effect generalizes beyond legacy CSI datasets.

Overall, this experiment shows that SDP fundamentally shifts evaluation from a stochastic, initialization-sensitive process to a controlled and reproducible benchmarking procedure. Importantly, this improvement is achieved without task-specific tuning, reinforcing that SDP’s primary contribution lies in standardizing the sensing pipeline rather than optimizing single-run accuracy.

\subsubsection{Ablation Study}

We conduct an ablation study to clarify why SDP improves benchmarking reliability, rather than where accuracy gains originate. The analysis follows a three-stage logic that mirrors the benchmarking pipeline itself: representation formation, training stability, and evaluation consistency.

We first examine representation-level effects through DFS spectrogram visualizations. As shown in Fig.~\ref{fig:ablation_spectrograms}, raw spectrograms across different datasets and actions are heavily contaminated by hardware-dependent noise and fragmented background energy. After SDP canonicalization, motion-induced patterns become sharper and more temporally coherent across tasks and environments. This result establishes that SDP fundamentally reshapes the CSI representation into a more physically interpretable and task-agnostic form, providing a stable input foundation for learning.

Building on this representation-level improvement, we next analyze seed-level training stability. Fig.~\ref{fig:ablation_seeds} reports the performance distribution over five random seeds for the full SDP pipeline and its ablated variants. Removing key components such as canonical grid construction or phase calibration reintroduces large performance variance and unstable convergence, approaching the behavior of native pipelines. In contrast, the full SDP configuration yields tightly clustered results, demonstrating that standardized preprocessing directly translates into reduced sensitivity to initialization during training.

Finally, we evaluate cross-run ranking consistency, which is critical for meaningful benchmark comparisons. Fig.~\ref{fig:ablation_rank} shows that the full SDP pipeline consistently maintains a top-ranked position across all seeds, whereas ablated variants exhibit frequent rank reversals. This indicates that, without SDP, relative method ordering becomes dependent on random initialization, undermining the statistical validity of comparative evaluation.

Together, these results form a coherent chain: SDP first regularizes the sensing representation, which stabilizes training dynamics, and ultimately ensures consistent ranking across runs. This confirms that SDP’s contribution lies not in task-specific optimization, but in enforcing a physically grounded and reproducible benchmarking interface.

\subsubsection{Computational Efficiency Analysis}

A potential concern with introducing a middleware layer is the additional computational overhead. We explicitly analyze the theoretical computational cost (FLOPs) of the SDP pipeline compared to native task-specific baselines. 

Taking the complex XRF55 dataset as a representative case, the native baseline typically concatenates raw CSI streams from multiple receivers into a high-dimensional input tensor (e.g., aggregating 3 receivers $\times$ 90 subcarriers $\approx$ 270 channels). This results in a heavy computational load of approximately 4.2 GFLOPs for a single inference window.

In contrast, SDP decomposes the input into three standardized canonical tensors (one per receiver path, each with a fixed dimension of $K=30$), which are processed efficiently. Although SDP introduces a preprocessing step, the resulting canonical representation significantly reduces the input redundancy. Our estimation shows that processing a single canonical tensor requires only 1.36 GFLOPs. For a 3-receiver setup, the total computational cost is $1.36 \times 3 \approx$ 4.08 GFLOPs.

\begin{table}[ht]
\centering
\caption{Algorithm-agnostic benchmarking}
\label{tab:algorithm_agnostic}
\begin{tblr}{
  width = \linewidth,
  colspec = {Q[373]Q[227]Q[131]Q[198]},
  cells = {c},
  hlines,
  vlines,
}
\textbf{Dataset/Task}            & \textbf{CNN / ResNet} & \textbf{BiLSTM} & {\textbf{Transformer}\\\textbf{(SDP)}} \\
\textbf{ElderAL-CSI (Act.+Loc.)} & 0.7987                & 0.8037          & 0.8235                                 \\
\textbf{GaitID (User ID)}        & 0.9829                & 0.9741          & 0.9843                                 
\end{tblr}
\end{table}

\begin{figure*}[hbp!]
    \centering
    \captionsetup{justification=raggedright}
    \newlength{\rwheight}
    \setlength{\rwheight}{7cm} 

    \begin{subfigure}[b]{0.495\textwidth}
        \centering
        \includegraphics[height=\rwheight, width=\linewidth, keepaspectratio]{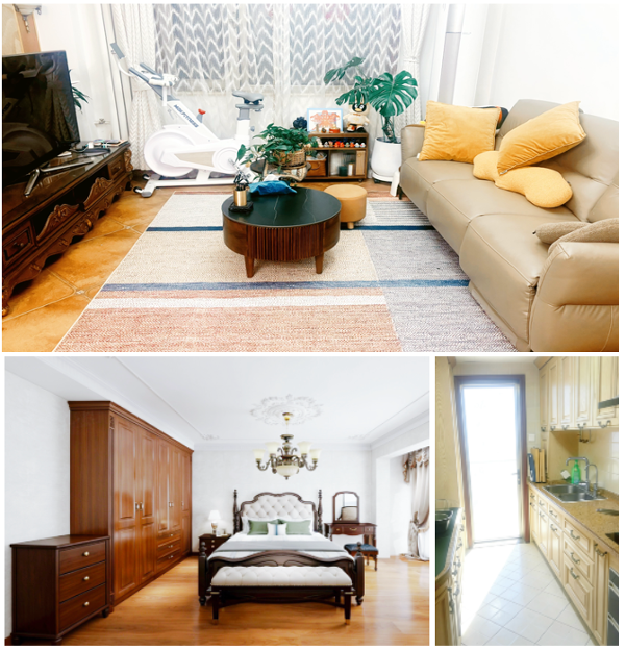} 
        \caption{Real-world Scenarios}
        \label{fig:real_world}
    \end{subfigure}
    \begin{subfigure}[b]{0.495\textwidth}
        \centering
        \includegraphics[height=\rwheight, width=\linewidth, keepaspectratio]{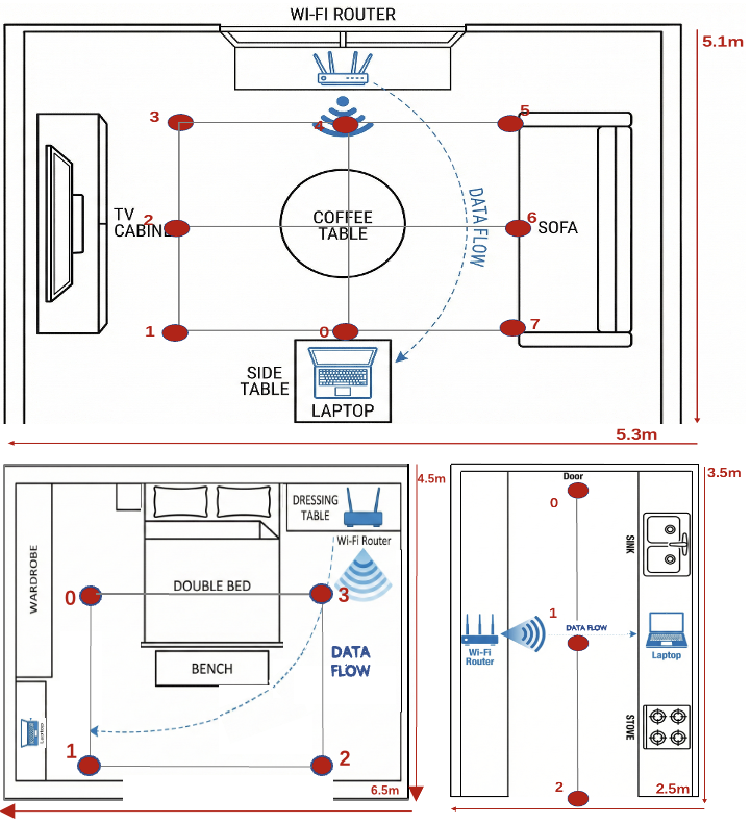} 
        \caption{Schematic Diagrams}
        \label{fig:schematic}
    \end{subfigure}

    \caption{Real-world deployment setup for protocol validation. (a) Photographs of three heterogeneous residential environments: a living room (primary), kitchen, and bedroom (auxiliary). (b) Corresponding floor plans illustrating distinct spatial configurations. This setup is used to qualitatively validate the interoperability and feasibility of SDP under realistic environmental conditions.}

    \label{fig:realworld_setup}
\end{figure*}

This result indicates that SDP is computationally neutral: it achieves standardization and stability without imposing additional latency or computational burden compared to native methods. The slight reduction in FLOPs implies that the cost of canonicalization is effectively offset by the efficiency gains from processing a cleaner, lower-dimensional manifold.

\subsubsection{Algorithm-Agnostic Benchmarking Capability}

To further demonstrate that SDP functions as a general-purpose benchmarking middleware rather than a model-specific optimization, we evaluate its compatibility with diverse backbone architectures under a unified configuration.

Specifically, three representative model families commonly used in wireless sensing are considered: a lightweight CNN-based architecture, a classical sequence-modeling backbone (BiLSTM), and the Transformer-based model adopted in the main benchmark. All models operate on the same canonical CSI tensor defined by SDP and are trained using an identical optimization protocol, without any architecture-specific tuning or task-dependent engineering. Experiments are conducted on two representative tasks with different characteristics, namely gait-based user identification (GaitID) and joint activity--location recognition on ElderAL-CSI.

The results indicate that SDP provides a stable and comparable evaluation interface across heterogeneous algorithms. The quantitative results are summarized in Table~\ref{tab:algorithm_agnostic}.
Although absolute performance varies across architectures, all models converge reliably and exhibit consistent behavior under the same SDP protocol.

It is important to emphasize that the objective of this experiment is not to identify the optimal backbone architecture, but to verify that SDP enables fair, stable, and reproducible comparison among different algorithms. By decoupling model design from dataset-specific preprocessing and training heuristics, SDP ensures that observed performance differences primarily reflect algorithmic capability rather than pipeline artifacts. This algorithm-agnostic property is essential for a benchmark intended for community use, as it allows future models to be seamlessly integrated into the canonical SDP interface and evaluated against existing results under a common and reproducible standard.

\subsection{Case Study: Real-world Deployment}

This case study complements the controlled benchmark results by examining whether the stability properties induced by SDP persist under uncontrolled residential data and continuous temporal dependencies.
Unlike laboratory benchmarks that rely on curated segmentation and near i.i.d.\ samples, this study introduces realistic domain shifts in spatial layout, multipath propagation, and human motion patterns.
Rather than establishing a new quantitative benchmark, the objective is to qualitatively stress-test whether SDP preserves stable and interpretable behavior when standard evaluation assumptions no longer hold.

\subsubsection{Heterogeneous Environments and Data Collection}

The deployment spans three representative residential environments, including a fully furnished living room, a compact kitchen, and a bedroom (Fig.~\ref{fig:realworld_setup}).
These scenarios differ substantially in room size, furniture density, and dominant motion ranges, thereby inducing heterogeneous multipath structures and occlusion patterns commonly encountered in practical in-home sensing.

\begin{figure*}[t!]
    \centering
    \captionsetup{justification=raggedright}

    \begin{subfigure}[b]{0.48\textwidth}
        \centering
        \includegraphics[width=\linewidth, height=5.0cm, keepaspectratio]{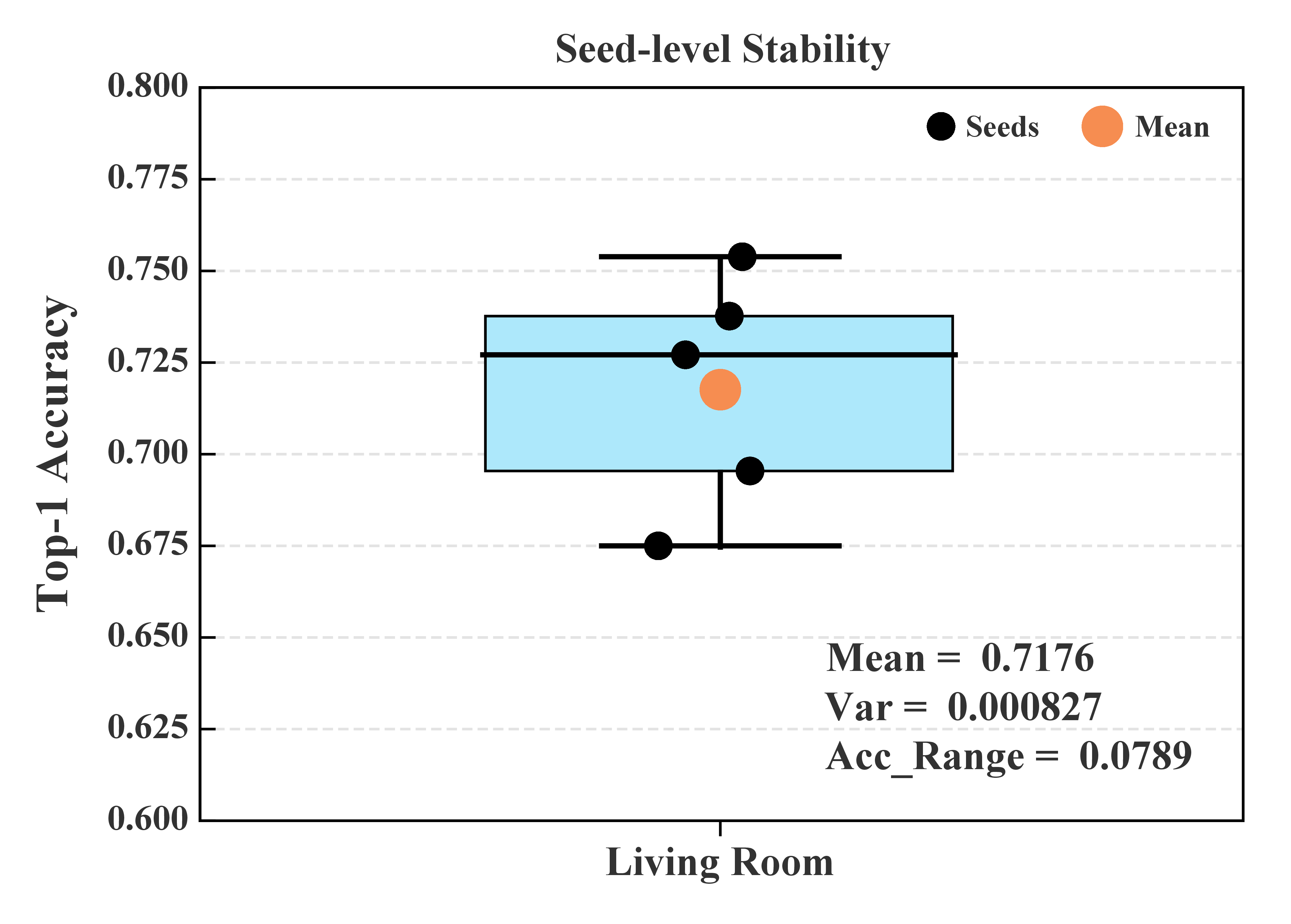}
        \caption{Seed-level accuracy distribution in the Living Room. Black markers denote individual random seeds, and the red marker indicates the mean.}
        \label{fig:stability_living}
    \end{subfigure}
    \hfill 
    \begin{subfigure}[b]{0.48\textwidth}
        \centering
        \includegraphics[width=\linewidth, height=5.0cm, keepaspectratio]{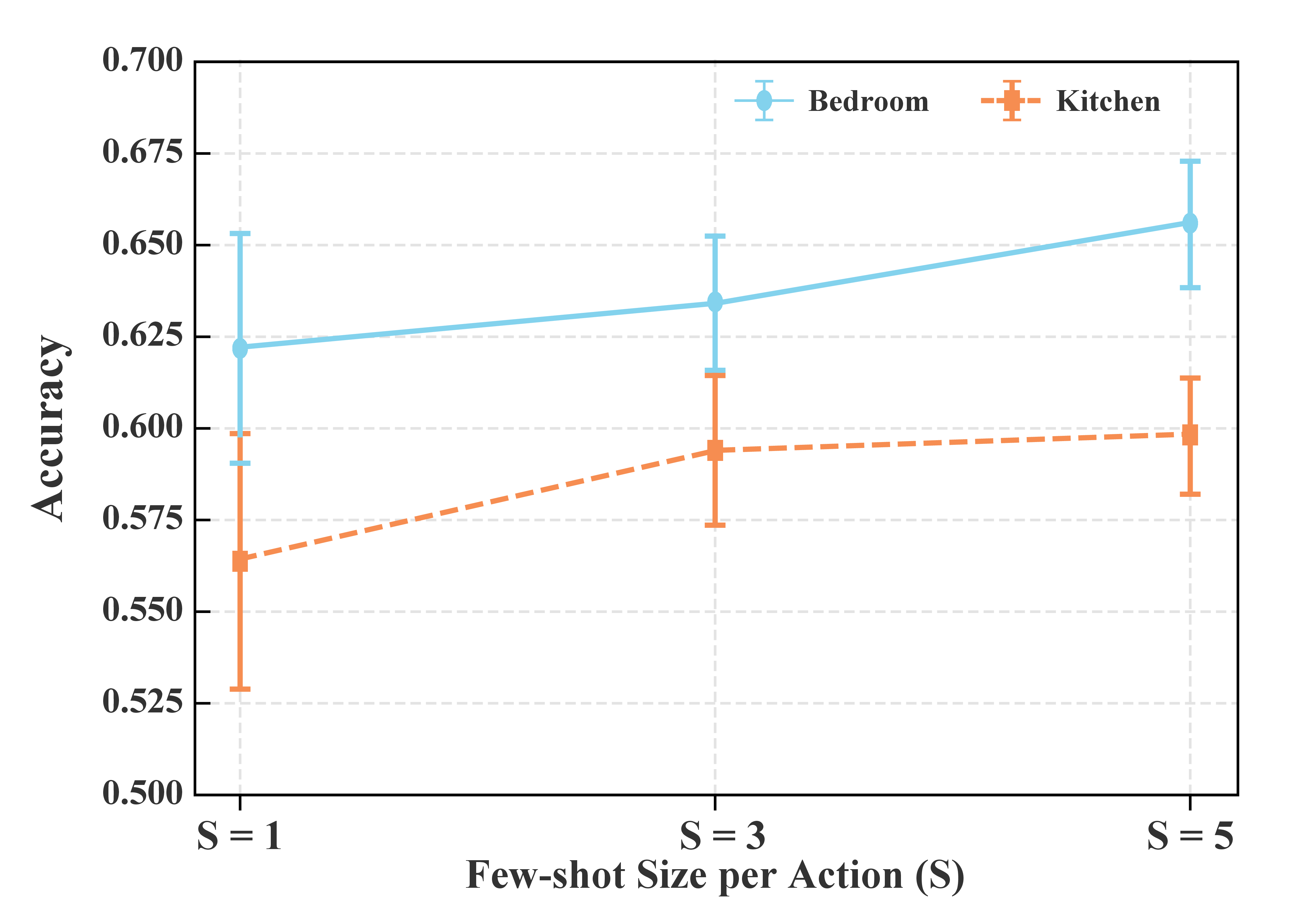}
        \caption{Few-shot adaptation accuracy in Bedroom and Kitchen scenarios. Results are reported for varying shot sizes ($S \in \{1, 3, 5\}$) with error bars.}
        \label{fig:fewshot_results}
    \end{subfigure}

    \vspace{2pt}

    \caption{Protocol-level robustness in real-world residential scenarios. (a) The compact distribution and bounded range in the Living Room suggest that SDP effectively suppresses pathological seed sensitivity, even with temporally redundant data. (b) Consistent performance trends in unseen Bedroom and Kitchen environments demonstrate the protocol's adaptability under limited supervision. Together, these results validate SDP's stability across heterogeneous spatial layouts and random initializations.}
    \label{fig:realworld_eval}
\end{figure*}

Among the three environments, the living room serves as the primary data collection scenario, while the kitchen and bedroom are used for few-shot cross-environment validation.
CSI data are collected using a commercial ZTE AX3000 Wi-Fi~6 router operating in the 5~GHz band.
The router is connected via Ethernet to a standard laptop, where CSI packets are received and recorded, subsequently processed using the SDP preprocessing pipeline.
Although similar COTS Wi-Fi~6 devices have appeared in recent datasets, this study differs in its end-to-end acquisition pipeline, canonical preprocessing, and multi-scenario evaluation setting, and therefore constitutes an independent real-world validation of the SDP protocol.

\subsubsection{Primary Living-room Dataset and Few-shot Cross-environment Validation}

The living room serves as the primary data collection and training environment.
Six spatial locations were selected to capture location-dependent channel variations.
Data were collected from three subjects performing five representative motion classes: Standing, Walking, Waving, Jumping, and Fall.
These actions span a broad range of motion dynamics, from quasi-static postures to abrupt high-energy events, while avoiding highly confusable transitional actions that are difficult to disambiguate under small-scale real-world collections.

For each subject--location--action combination, data collection was repeated 15 times, with each trial lasting 5~s (approximately 500 CSI frames at a nominal sampling rate of 100~Hz).
All measurements were processed using the same SDP pipeline described in Section~\ref{section3}, including deterministic STO/CFO sanitization, canonical frequency projection, and fixed-length temporal segmentation into the Canonical CSI Tensor
$\mathcal{X} \in \mathbb{C}^{A \times K \times T}$.
This living-room dataset constitutes the sole training source for the case study.

To examine environment robustness under limited supervision, additional data were collected in the kitchen and bedroom scenarios.
Rather than constructing fully balanced datasets, these environments were intentionally used for few-shot cross-environment validation.
Specifically, for each action, only a small number of labeled samples (with few-shot sizes $S\in\{1,3,5\}$ per action) were collected per subject and location, resulting in lightweight adaptation sets.
These samples were used for lightweight fine-tuning or direct evaluation on models trained solely in the living room.
This setting reflects realistic deployment constraints, where collecting large-scale labeled data in every new environment is impractical.

\begin{figure*}[t!]
    \centering
    \captionsetup{justification=raggedright}

    \newlength{\topheight}
    \setlength{\topheight}{4.5cm} 

    \newlength{\bottomheight}
    \setlength{\bottomheight}{3.0cm}

    \begin{subfigure}[b]{\linewidth}
        \centering
        \includegraphics[width=\linewidth, height=\topheight, keepaspectratio]{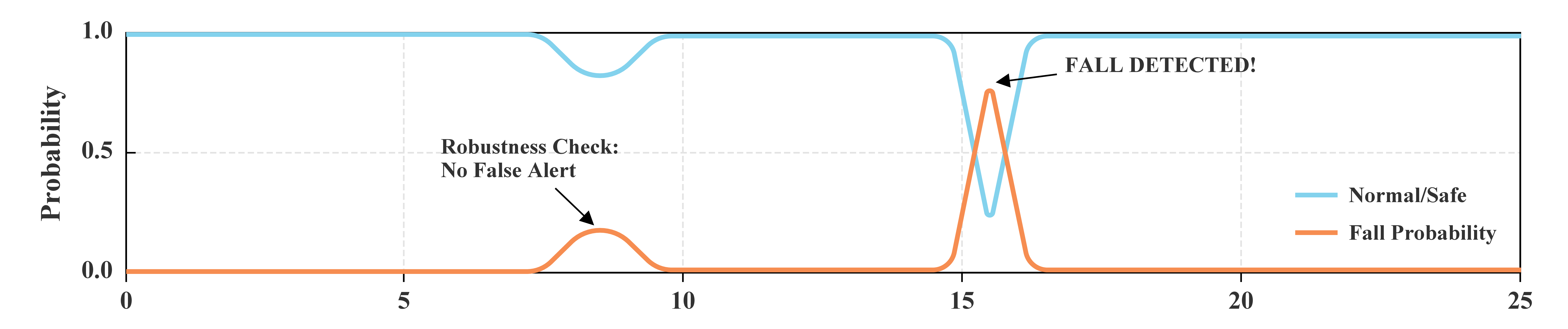}
        \caption{Continuous CSI stream processing after SDP canonicalization}
        \label{fig:continuous_streams}
    \end{subfigure}

    \vspace{1pt} 

    \begin{subfigure}[b]{0.25\linewidth} 
        \centering
        \includegraphics[width=\linewidth, height=\bottomheight, keepaspectratio]{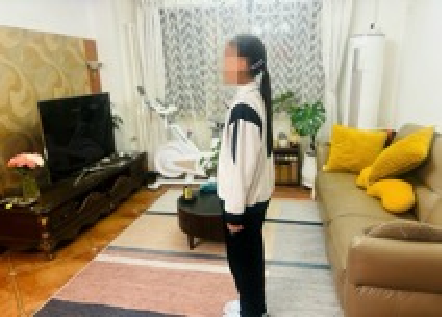}
        \caption{Standing}
        \label{fig:stand}
    \end{subfigure}
    \hspace{0.5cm} 
    \begin{subfigure}[b]{0.25\linewidth} 
        \centering
        \includegraphics[width=\linewidth, height=\bottomheight, keepaspectratio]{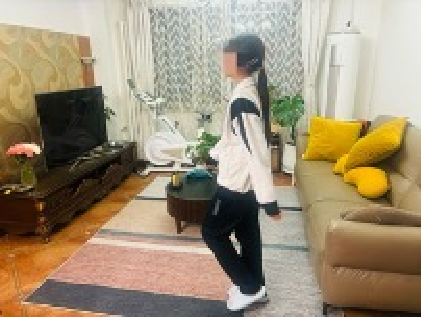}
        \caption{Walking}
        \label{fig:walk}
    \end{subfigure}
    \hspace{0.5cm} 
    \begin{subfigure}[b]{0.25\linewidth} 
        \centering
        \includegraphics[width=\linewidth, height=\bottomheight, keepaspectratio]{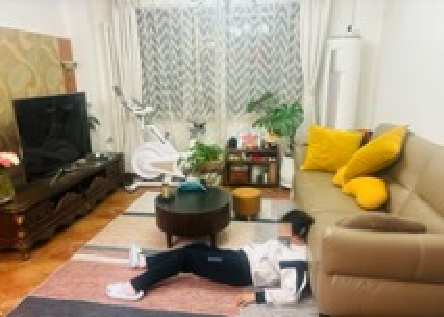}
        \caption{Fall}
        \label{fig:fall}
    \end{subfigure}

   \caption{Continuous-stream inference illustration in a residential environment. (a) Example of a continuous CSI stream after SDP canonical preprocessing, used for offline stream-level inference without oracle segmentation. (b--d) Representative CSI spectrograms corresponding to standing, walking, and falling activities, highlighting distinct time--frequency patterns preserved by the canonical representation.}
    \label{fig:continuous_streams_overview}
\end{figure*}

\subsubsection{Results and Stability Analysis}

We first examine seed-level behavior in the primary Living Room environment, where the model is trained solely on canonicalized CSI samples generated by the SDP pipeline. As shown in Fig.~\ref{fig:stability_living}, the Top-1 accuracy across five random seeds concentrates within a relatively narrow range, with an average accuracy of 0.7176 and a variance of $8.27\times10^{-4}$.
The maximum accuracy difference across seeds remains below 0.08, indicating that no severe performance collapse or seed-dependent instability is observed under repeated training. Rather than producing highly volatile outcomes across random initializations, the learned models exhibit consistent convergence behavior when operating on SDP-canonicalized CSI representations.
This observation suggests that the protocol-level preprocessing mitigates excessive sensitivity to correlated, non-i.i.d.\ samples, which are unavoidable in practical residential sensing scenarios.

We further analyze cross-environment robustness through few-shot adaptation in the Bedroom and Kitchen scenarios. Fig.~\ref{fig:fewshot_results} reports the mean accuracy and standard deviation over five random seeds for few-shot sizes $S \in \{1, 3, 5\}$.
Although these environments differ substantially from the training scenario in terms of room geometry and multipath structure, performance increases smoothly as limited supervision is introduced, without observable variance amplification across seeds.
The adaptation trends remain consistent across both environments, indicating predictable behavior under constrained supervision rather than abrupt or unstable responses.

Overall, these results do not aim to demonstrate peak recognition accuracy in real-world settings.
Instead, they indicate that the stability-oriented properties observed in controlled benchmark analyses persist under realistic residential data, where temporal redundancy and environment shifts are unavoidable.
This alignment supports the role of SDP as a protocol-level middleware that yields stable and interpretable evaluation behavior beyond idealized i.i.d.\ assumptions.

\subsubsection{Continuous-stream Canonical Inference}

Beyond window-level evaluation, we further analyze SDP under continuous CSI streams without oracle action segmentation. In this setting, CSI packets are first canonicalized according to the SDP protocol and then sequentially fed into the unified sensing model for offline stream-level inference.

We construct continuous sequences of approximately 25~s that contain multiple daily activities as well as a safety-critical fall event.
For each sequence, the model outputs a per-frame probability for the \emph{Fall} class, denoted as $p_{\text{fall}}(t)$.
To summarize stream-level behavior, we adopt the peak fall probability
\begin{equation}
p_{\text{fall}}^{\max} = \max_{t \in [0,25\text{s}]} p_{\text{fall}}(t),
\end{equation}
which reflects the model’s confidence in detecting a fall within a continuous sequence.
For reporting purposes, a fall response is counted when $p_{\text{fall}}^{\max}$ exceeds a fixed confidence threshold $\tau$, where $\tau$ is kept constant across all trials and environments.
This definition is used solely as a protocol-level indicator and does not correspond to a tuned event-detection rule.

Across 10 independent continuous-sequence trials, the fall event is indicated in 8 trials according to the above protocol-level criterion based on $p_{\text{fall}}^{\max}$ and the corresponding ground-truth timestamps.
Fig.~\ref{fig:continuous_streams_overview} illustrates a representative example.
The canonicalized inference output maintains low fall probability during normal activities while producing a sharp confidence peak around the fall moment, indicating reduced false-alarm sensitivity and temporally localized responses under canonical preprocessing.

Taken together, this case study does not aim to demonstrate a deployable sensing system, but to examine whether SDP preserves stable and interpretable behavior when standard i.i.d.\ assumptions are violated. The observed consistency between benchmark stability analysis and continuous-stream inference further supports SDP’s role as a protocol-level middleware for reproducible wireless sensing evaluation.

\section{Conclusion}\label{section6}

This paper presented the SDP, a protocol-level abstraction and unified benchmark designed to address the long-standing reproducibility and comparability challenges in wireless sensing research.
Unlike prior works that primarily focus on task-specific models or isolated performance improvements, SDP targets the root cause of instability in wireless sensing systems by enforcing deterministic physical-layer sanitization, canonical tensor construction, and standardized learning and evaluation procedures.

Through extensive experiments across multiple representative sensing tasks and heterogeneous datasets, we demonstrated that SDP consistently achieves competitive accuracy while dramatically reducing performance variance across random initializations. In particular, SDP yields order-of-magnitude reductions in inter-seed variance for complex activity recognition tasks. These results suggest that a significant portion of the observed instability in wireless sensing stems from protocol-level ambiguity rather than inherent limitations of learning-based approaches.

Beyond controlled benchmarks, a real-world deployment using COTS Wi-Fi hardware further validated the hardware-agnostic nature and practical feasibility of the proposed protocol. Despite pronounced domain shifts in sensing hardware, spatial layout, and multipath conditions, the SDP pipeline enabled seamless canonicalization and reliable inference without modifying the unified model. This case study highlights SDP's potential as a protocol middleware applicable beyond dataset- or device-specific solutions.

Overall, SDP provides a principled foundation for reproducible wireless sensing research by decoupling learning performance from hardware-specific artifacts.
By establishing a common protocol and benchmark, SDP enables fair comparison across sensing tasks, devices, and experimental setups, and paves the way toward more reliable and interpretable sensing systems.

Future work will explore extending SDP to broader sensing modalities and more challenging deployment scenarios, including large-scale multi-device environments, cross-domain generalization, and resource-constrained edge platforms. We believe that protocol-driven standardization, as advocated by SDP, will play a critical role in advancing wireless sensing from ad hoc experimentation toward a mature and trustworthy engineering discipline.


%

\ifCLASSOPTIONcaptionsoff
  \newpage
\fi

{
\small
\bibliographystyle{IEEEtran}
\bibliography{bibtex/bib/IEEEabrv,bibtex/bib/IEEEexample}
}
\end{document}